\documentclass[12pt]{iopart}
\usepackage{iopams,graphicx}

\usepackage{lmodern}
\usepackage[T1]{fontenc}
\usepackage{bbold}
\graphicspath{{graphics/}}
\usepackage{xcolor}

\usepackage{soul}
\usepackage[mathscr]{euscript}
\usepackage{cite}
\usepackage[super]{nth}
\usepackage{float}  
\usepackage{comment}                

\begin{document}                

\title{Intermittent resetting potentials}

\author{Gabriel Mercado-V\'asquez$^1$, Denis Boyer$^1$,	Satya N Majumdar$^2$ and Gr\'egory Schehr$^2$}

\address{$^1$ Instituto de F\'isica, Universidad Nacional Aut\'onoma de M\'exico, Mexico City 04510, Mexico}
\address{$^2$ LPTMS, CNRS, Univ. Paris-Sud, Universit\'e Paris-Saclay, 91405 Orsay, France}

%


\begin{abstract}
We study the non-equilibrium steady states and first passage properties of a Brownian particle with position $X$ subject to an external confining potential of the form $V(X)=\mu|X|$, and that is switched on and off stochastically. Applying the potential intermittently generates a physically realistic diffusion process with stochastic resetting toward the origin, a topic which has recently attracted a considerable  interest in a variety of theoretical contexts but has remained challenging to implement in lab experiments. The present system exhibits rich features, not observed in previous resetting models. The mean time needed by a particle starting from the potential minimum to reach an absorbing target located at a certain distance can be minimized with respect to the switch-on and switch-off rates. The optimal rates undergo continuous or discontinuous transitions as the potential strength $\mu$ is varied across non-trivial values. A discontinuous transition with metastable behavior is also observed for the optimal strength at fixed rates.
\end{abstract}

\noindent{\it Keywords\/}: Brownian motion, driven diffusive systems, stationary states, stochastic searches


\maketitle

\section{Introduction}

When searching unsuccessfully for a hidden item, after some time it may be convenient to return toward the starting point and resume exploration afresh from there. The idea that the completion of a task or the encounter with a goal can be expedited by restart (or resetting) is often used in computer science and physics, for instance for addressing hard problems of combinatorial optimization \cite{montanari2002optimizing}. More recently, diffusive processes with stochastic resetting (see \cite{evans2020stochastic}
for a recent review) have been found relevant to a broad range of phenomena, such as enzymatic reactions \cite{Reuveni4391}
, adaptive evolution in genetics \cite{bittihn2017gene}, foraging ecology \cite{boyer2014random,giuggioli2019comparison}, active transport in living cells \cite{bressloff2020modeling}, or power management problems \cite{de2020first}. 

Over the past decade, one-dimensional models have been instrumental for understanding the effects of resetting on stochastic processes, in particular the emergence of non-equilibrium steady states (NESS) \cite{manrubia1999stochastic,evans2011diffusion,Evans_2018}, their relaxation dynamics \cite{majumdar2015dynamical}, as well as the consequences on first passage properties\cite{evans2020stochastic,Evans_2018}. The mean first passage time (MFPT) to a target, which is infinite for a purely diffusive process in an unbounded domain, becomes finite under resetting and can be minimized with respect to the resetting rate \cite{evans2011diffusion}. This property has sparked much interest and a body of theoretical results is now available on standard diffusion models under stochastic resetting \cite{evans2011diffusionb,montero2013monotonic,evans2014diffusion,kusmierz2014first,kusmierz2015optimal,christou2015diffusion,reuveni2016optimal,riascos2020random} or under more general resetting schemes \cite{eule2016non,nagar2016diffusion,pal2016diffusion,boyer2017long, falcon2017localization,chechkin2018random}.

Meanwhile, very few experiments have been conducted to verify theoretical predictions or uncover new features associated to resetting. The paradigmatic problem of the one-dimensional Brownian motion with resetting to the origin \cite{evans2011diffusion} has been implemented with silica micro-spheres manipulated by optical tweezers \cite{besga2020optimal,tal2020experimental}. These experimental realizations pose several challenges, as the original model assumes that the particle is always relocated exactly at the same position, and in zero time. The theory had to be modified to consider a distribution of restart locations of finite width \cite{besga2020optimal}, leading to the observation of interesting metastability effects for the MFPT. Models with non-instantaneous resetting have also been proposed recently \cite{maso2019transport,pal2019time,bodrova2020resetting,bodrova2020two,pal2019home,gupta2020stochastic} and compared to experiments that used different types of return motion, {\it e.g.} at a constant speed or constant time \cite{tal2020experimental}.

In this study, we address a generic and physically realistic resetting problem, which in principle would not require 
to track a particle or return it to the origin in a controlled deterministic way. The method consists in using an external trap, namely, a symmetric confining potential with a single minimum, to attract the particle toward the origin (see also \cite{gupta2020stochastic}). 
Consider a Brownian particle in one dimension with diffusion constant $D$ and friction coefficient 
set to unity,
driven by the action of an intermittent potential. The state of the potential is described by a time dependent binary variable $\sigma(t)$, where $\sigma(t)=0$ means that the potential $V(X)$ is switched off, and $\sigma(t)=1$, that it is applied. 
The two-state process $\sigma(t)$ is characterized by constant transition rates, $R_0$ (for the transition $0\rightarrow1$) and $R_1$ (for $1\rightarrow0$). 
Here, we wish to elucidate whether an intermittent potential can generate non-equilibrium steady states for the particle probability density, and whether it may facilitate a target search, as in genuine resetting models.

\begin{figure}[t]
    \centering
    \includegraphics[width=1\linewidth]{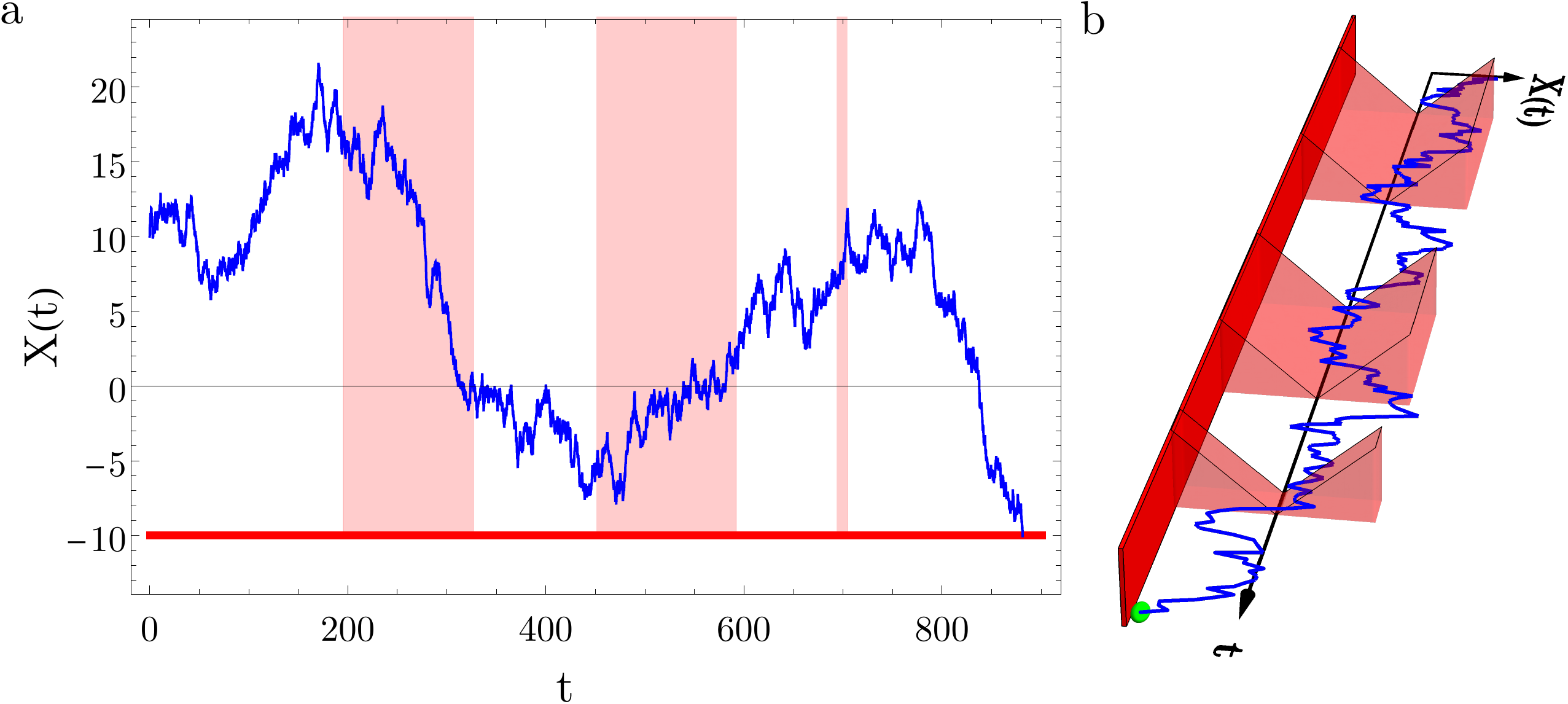}
    \caption{a) Trajectory of a diffusive particle with diffusion constant $D=1$, in an intermittent resetting potential $V(X)=\mu|X|$ with $\mu=0.1$. The shaded zones represent the time intervals when the potential is turned on ($R_0=R_1=0.005$). An absorbing boundary is placed at $-10$. b) $3D$ view of a particle trajectory and a potential with $\mu=1$. 
    }
    \label{fig:graphic1}
\end{figure}

Figure \ref{fig:graphic1} depicts some trajectories in the presence of an absorbing boundary (target) placed at a distance $L$ from the origin, on the negative side.
Free diffusion is interspersed with periods of potential reset, during which the particle is attracted toward the origin. The motion is similar to diffusion with a non-instantaneous resetting protocol for the particle position, with three important differences compared to recent studies on this subject \cite{maso2019transport,pal2019time,bodrova2020resetting,bodrova2020two,pal2019home,gupta2020stochastic}: 1) the return toward the origin is not deterministic, owing to fluctuations; 2) the particle is not necessarily at the potential minimum $X=0$ when the potential is switched off, as the dynamics of $\sigma(t)$ is independent of the particle position; 3) the target at $X=-L$ is always detectable, {\it i.e.}, it can be found when the potential state is either $0$ or $1$. Therefore the search process is not suspended during the \lq\lq on\rq\rq$\ $ phase, as often assumed. The features 1)--3) in combination lead to novel phenomena, as we will see below.

The evolution of the particle position $X(t)$ in the potential $\sigma(t)V(X)$ is given by the over-damped Langevin equation:
\begin{equation}
    \frac{dX(t)}{dt}= -\sigma(t)V^{\prime}[X(t)]+\xi(t),
    \label{ruleX1}
\end{equation}
with $\xi(t)$ a Gaussian white noise of zero mean and correlations $\langle \xi(t)\xi(t')\rangle=2D\delta(t-t')$. In this study, for analytical convenience we work with linear potentials, $V(X)=\mu|X|$ where $\mu>0$. We expect other confining potentials to yield qualitatively similar findings. 

There are several limiting cases that ought to be mentioned. If $R_1=0$ and $\sigma(t=0)=1$, the potential is stationary and, in the unbounded domain, $X(t)$ follows the Boltzmann-Gibbs equilibrium distribution $\frac{\mu}{2D}\exp[-\mu|X|/D]$ at large times. Another limit is that of infinite potential strength, more specifically $\mu L/D\rightarrow\infty$, which corresponds to a nearly instantaneous resetting to the origin at rate $R_0$, followed by a refractory time (exponentially distributed and of mean $1/R_1$) during which the particle stays immobile \cite{evans2018effects}. If $\mu L/D=\infty$ and $R_1\rightarrow\infty$, one recovers the standard problem of diffusion with stochastic resetting to the origin at rate $R_0$, where the particle is released immediately after resetting \cite{evans2011diffusion}.

The paper is organized as follows. In Section \ref{sec:sum}, we present a summary of the results. The asymptotic particle density is derived in Section \ref{sec:dens}. Section \ref{sec:mfpt} is devoted to the MFPTs. Therein, a heuristic calculation is presented (Section \ref{sec:naive}), followed by an exact perturbative theory (Sections \ref{sec:perturb} and  \ref{sec:t0}). In Section \ref{sec:meta}, the metastable behavior of the MFPTs with respect to the potential strength is analysed, before the Conclusions. Technical details on the exact solution of the MFPTs and on the perturbative calculations have been left in \ref{sec:exact} and \ref{appendixt2}, respectively.

\section{Summary of the results}\label{sec:sum}

For the convenience of the reader we provide a summary of the main results, so that they can be understood without going through the details that are laid out in later sections.
We introduce the dimensionless space and time variables $x=X/L$ and $t/(L^2/D)$ (which we renote as $t$). The problem is fully described by three dimensionless parameters:
\begin{eqnarray}
r_0&=&R_0L^2/D, \label{adimr0}\\ 
r_1&=&R_1L^2/D, \label{adimr1}\\
\gamma&=&\mu L/D, \label{adimgamma}
\end{eqnarray}
namely, the re-scaled \lq\lq on\rq\rq$\ $ and \lq\lq off\rq\rq $\ $ rates, and the re-scaled potential strength, respectively. 

\subsection{Stationary density}

Similarly to many resetting processes, on the unbounded infinite line the problem admits a non-equilibrium stationary density $p(x)$ for the particle position (irrespective of the potential state). This NESS exits for any $\gamma>0$, $r_0>0$ and $r_1>0$, and it is the sum of two exponentials:
\begin{equation}
p(x)=a_1e^{-\lambda_1 |x|}+a_3e^{-\lambda_3 |x|},
\end{equation}
where $\lambda_1$ and $\lambda_3$ are the two positive roots of the equation $\lambda^3-\lambda^2 \gamma -\lambda(r_1+r_0)+r_0\gamma=0$, and are given by Eqs. (\ref{lambdas})-(\ref{theta}).
The prefactors $a_1$ and $a_3$ can be read off from Eq. (\ref{ptotal}).

In the large potential strength limit, $\gamma\gg\sqrt{r_1}$ and $\gamma\gg r_1/\sqrt{r_0}$, this density takes the simpler form:
\begin{equation}
     p(x)\simeq \frac{ r_0}{r_0+r_1}\left( \frac{\gamma}{2}e^{-\gamma |x|}\right)+\frac{r_1}{r_0+r_1}\left(\frac{\sqrt{r_0}}{2} e^{-\sqrt{r_0} |x|}\right),\label{eq:denslargegamma}
\end{equation}
which is the average between the Boltzmann-Gibbs distribution in the presence of the potential and the NESS of free diffusion with instantaneous resetting (at rate $r_0$). 

\subsection{MFPT: optimal protocols}

We calculate exactly 
in \ref{sec:exact} 
two MFPTs (re-scaled by $L^2/D$) for a target located at the dimensionless position $-1$. They are denoted by $t_0(x)$ and $t_1(x)$, where $x$ represents the initial position of the particle and the subscript $0$ (respectively $1$) refers to the initial potential state $\sigma(t=0)=0$ (respectively $\sigma(t=0)=1$). These mean times are finite as soon as $\gamma>0$ and $r_0>0$, and exhibit rather unexpected behaviours.
We consider a particle starting at the minimum of the potential ($x=0$), which is representative of experiments with optical tweezers. The $x$ dependence will be dropped when it is clear from the context.

Let us first fix the strength $\gamma$ and vary the rates, seeking to minimize the MFPTs $t_1$ and $t_0$ in the $(r_0,r_1)$-plane. We start with $t_1$ (the potential is \lq\lq on\rq\rq$\ $ at $t=0$) and define
\begin{equation}
t_1^*(\gamma)=\displaystyle\min_{r_0,r_1} t_1(\gamma,r_0,r_1),
\end{equation}
while $r_0^*(\gamma)$ and $r_1^*(\gamma)$ denote the corresponding optimal rates.  
We show that the optimal parameter $r_1^*(\gamma)$ exhibits a second-order transition at a non-trivial critical potential strength
\begin{equation}
\gamma_c=1.228780... 
\end{equation}
By \lq\lq second-order\rq\rq, we mean that $r_1^*(\gamma)$ 
goes to zero continuously at $\gamma=\gamma_c$. 
(Conversely, in a first-order transition, $r_1^*$ or $r_0^*$ would exhibit discontinuous jumps.)

\begin{figure}[t]
\centering
\includegraphics[width=\textwidth]{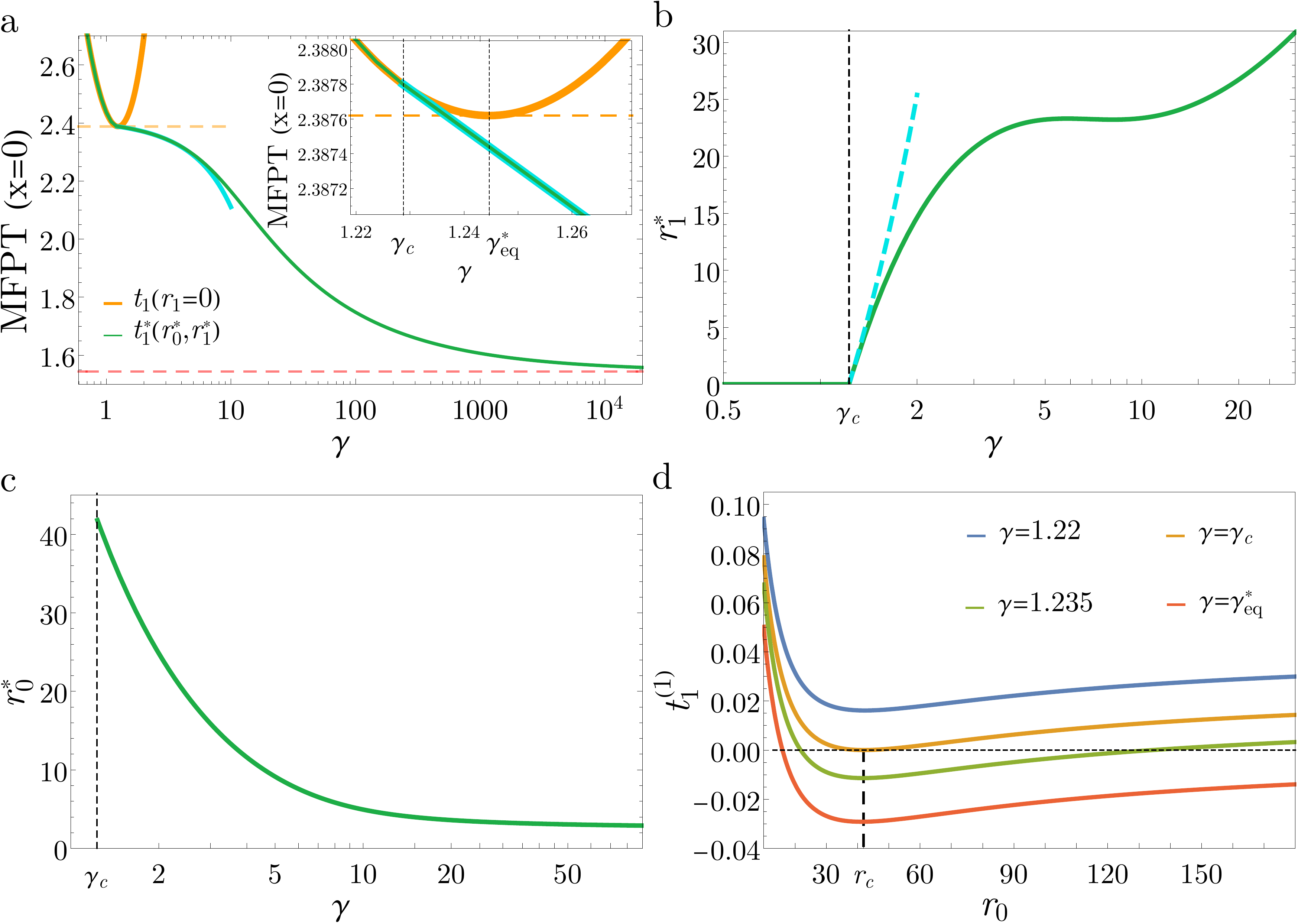}
\caption{Searches starting from $x=0$ and with the potential on at $t=0$.  (a) Minimal MFPT $t_1^*$ as a function of $\gamma$. The green line is obtained from numerical minimization of the exact solution with respect to $(r_0,r_1)$. The orange line represents $t_1$ for a particle in a steady potential ($r_1=0$), and the aqua line the analytical expression (\ref{t1optcrit}) for $\gamma>\gamma_c=1.2287...$. The lower horizontal dotted line is the limit $\gamma\rightarrow\infty$, corresponding to the optimized MFPT with instantaneous resetting. Inset: zoom of the transition region. (b) and (c): optimal rates $r_1
^*(\gamma)$ and $r_0^*(\gamma)$. (d) Coefficient of the first correction in the series expansion of $t_1$ near $r_1=0$ (the \lq\lq dispertion relation\rq\rq) as a function of $r_0$, for various $\gamma$ near $\gamma_c$. 
}
\label{fig:t1optgamma}
\end{figure} 

For $\gamma\le\gamma_c$, the resetting protocol that minimizes $t_1$ consists in keeping the potential always turned on, or
\begin{equation}
    r_1^*(\gamma)=0,
\end{equation}
while $r_0$ is irrelevant. The optimal MFPT is thus given in this regime by the Kramers' time of the equilibrium dynamics, denoted here as $t_{eq}(\gamma)$ [obtained by solving Eqs. (\ref{Tsystem2}) and (\ref{Tsystem2minus}) with $r_1=0$, see also \cite{evans2013optimal}]:
\begin{equation}\label{t1eq}
t_1^*(\gamma)=t_1(\gamma,r_0,r_1=0)\equiv t_{eq}(\gamma)=\frac{2}{\gamma^2}(e^{\gamma}-1)-\frac{1}{\gamma}.
\end{equation}
This function is depicted in Fig. \ref{fig:t1optgamma}a.

For $\gamma>\gamma_c$, in contrast, the MFPT
$t_1$ is minimized through a non-trivial resetting protocol.
Above the critical strength, the optimal switch-off rate is non-zero and increases rapidly with $\gamma$:
\begin{equation}\label{r1opt1gammac}
    r_1^*(\gamma)= 32.91301557...(\gamma-\gamma_c) + o\left(\gamma-\gamma_c\right),
\end{equation}
as shown in Fig. \ref{fig:t1optgamma}b. Meanwhile, the optimal switch-on rate $r_0^*(\gamma)$ decreases with $\gamma$ (Fig. \ref{fig:t1optgamma}c). Right at the transition, it is finite and surprisingly large:
\begin{equation}\label{r0opt1gammac}
    r_0^*(\gamma_c)=41.969027...
\end{equation}
Hence, as the potential strength increases, the most efficient searches are achieved by shortening the periods with the potential and extending the phases of free diffusion. The resulting MFPT also decreases with $\gamma$ (Fig. \ref{fig:t1optgamma}a). In the limit $\gamma\rightarrow\infty$, we see that $r_1^*\rightarrow\infty$ and $r_0^*\rightarrow 2.5396...$,  {\it i.e.}, we recover the optimal resetting rate for the well-known diffusion problem with instantaneous stochastic resetting \cite{evans2011diffusion}.

Above $\gamma_c$, the optimal searches are more efficient than in equilibrium, {\it i.e.}, $t_1^*(\gamma)<t_{eq}(\gamma)$. In the vicinity of $\gamma_c$,
\begin{equation}\label{t1optcrit}
\fl     t^*_1(\gamma)= t_{eq}(\gamma_c)-0.0225432...\left(\gamma-\gamma_c\right)-0.00103404...\left(\gamma-\gamma_c\right)^2+o\left((\gamma-\gamma_c)^2\right),
\end{equation}
a relation which remains accurate up to $\gamma-\gamma_c\sim 6$ in Fig. \ref{fig:t1optgamma}a.
Equation (\ref{t1optcrit}) is the numerical evaluation of Eq. (\ref{eq2ndt1}), which shows that the optimal MFPT  $t_1^*(\gamma)$ and its first derivative are continuous across the transition, whereas its second derivative is discontinuous.

We comment on an important point: the function $t_{eq}(\gamma)$  is non-monotonous and admits a minimum at $\gamma_{eq}^*=1.244678...$, a value which is above but very close to $\gamma_c$ (see the orange line in the inset of Fig. \ref{fig:t1optgamma}a). Hence,  unlike the prediction of a naive argument (Section \ref{sec:naive}), the optimal MFPT $t_1^*(\gamma)$ departs from $t_{eq}(\gamma)$ not at $\gamma_{eq}^*$ but slightly before, when it reaches the value $t_{eq}(\gamma_c)=2.387797...$, which is strikingly close to but above $t_{eq}(\gamma_{eq}^*)=2.387619...$

To unveil the mechanism of this unexpected transition, we developed a perturbative theory near $r_1=0$ (Section \ref{sec:perturb}) where the MFPT $t_1$ is expanded in powers of $r_1/r_0$:
\begin{equation}\label{landausum}
t_1(\gamma,r_0,r_1)=t_{eq}(\gamma)+\left(\frac{r_1}{r_0}\right)t_1^{(1)}(\gamma,r_0)+\left(\frac{r_1}{r_0}\right)^2 t_1^{(2)}(\gamma,r_0)+...
\end{equation}
The coefficient $t_1^{(1)}(\gamma,r_0)$ of the first correction is given by Eq. (\ref{t1approx1}), which is one of our main results. It is non-monotonous with $r_0$ and displayed in Figure \ref{fig:t1optgamma}d for various $\gamma$. A phase transition occurs when this function changes sign. The critical parameter $\gamma_c$ is such that for any $\gamma<\gamma_c$, the coefficient $t_1^{(1)}$ remains positive for all $r_0$: switching off occasionally a potential of subcritical strength (by setting $r_1\ne0$ but small) will always incur an increase of the MFPT compared to the case $r_1=0$. For any $\gamma>\gamma_c$, however, there exists a non-trivial interval of values of $r_0$ such that $t_1^{(1)}(\gamma,r_0)<0$. It is thus possible in this case to choose a finite rate $r_0$ that will produce a \emph{decrease} of the MFPT when $r_1$ is set to a small value. The critical point $\gamma_c$ is determined from the marginal curve $t_1^{(1)}(\gamma_c,r_0)$,
which vanishes at a single point $r_0=r_c$. The marginal rate is $r_c=41.969027...$, see Fig. \ref{fig:t1optgamma}d.

Meanwhile, the coefficient $t_1^{(2)}$ of the second order term of the expansion (\ref{landausum}) is given by Eq. (\ref{t12}) and is positive at $(\gamma_c,r_c)$. Therefore Eq. (\ref{landausum}) can be minimized with respect to $r_1$ close to the critical point and yields Eq. (\ref{r1opt1gammac}). One also deduces the optimal MFPT $t_1^*(\gamma)$ as a series expansion in $\gamma-\gamma_c>0$ (see Eq. (\ref{eq2ndt1}) or (\ref{t1optcrit}) above), as detailed in Section \ref{sec:perturb}. 

\begin{figure}[t]
\centering
\includegraphics[width=\textwidth]{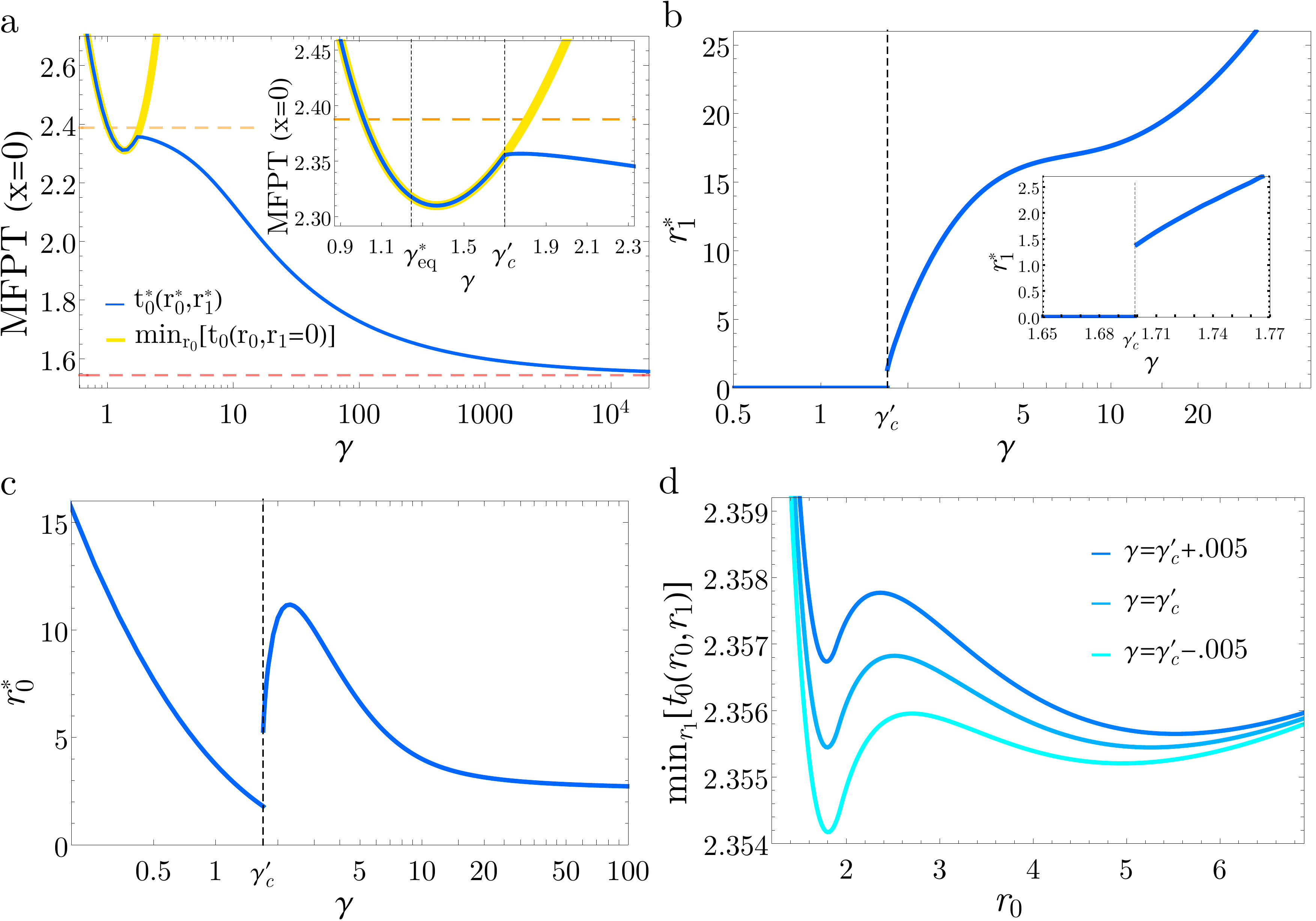}
\caption{Searches starting with a free diffusion phase. (a) Minimal MFPT $t_0^*$ as a function of $\gamma$. (b) Optimal rate $r_1^*(\gamma)$, with a zoom in the Inset. (c) Optimal rate $r_0^*(\gamma)$.  The yellow line in (a) and the region $\gamma<\gamma_c^{\prime}$ in (c) follow from the minimization of Eq. (\ref{fgammar0}). (d) $t_0$ exhibits metastability near $\gamma_c^{\prime}$: for instance, $\min_{r_1} t_0(\gamma,r_0,r_1)$ has local and global minima in $r_0$, resulting in a discontinuous $r_0^*$.
}
\label{fig:t0optgamma}
\end{figure}

It turns out that the properties of $t_0$, the MFPT of the particle starting with a free diffusion phase, are quite different. Remarkably, to the optimal time 
\begin{equation}
t_0^*(\gamma)=\displaystyle\min_{r_0,r_1} t_0(\gamma,r_0,r_1)
\end{equation}
is associated a \lq\lq first-order\rq\rq$\ $ transition at another potential strength $\gamma_c^\prime=1.698768...>\gamma_c$, see Figure \ref{fig:t0optgamma}a. In this case, the optimal rates $r_0^*$ and $r_1^*$ are discontinuous at $\gamma_c^\prime$ (Figs. \ref{fig:t0optgamma}b and \ref{fig:t0optgamma}c). In addition, the optimal MFPT $t_0^*(\gamma)$ slightly outperforms $t_1^*(\gamma)$ for any potential strength.

For $\gamma<\gamma_c^\prime$, one finds $r_1^*(\gamma)=0$ and $r_0^*(\gamma)>0$. Hence, to achieve a minimal MFPT with a low strength, the potential must be switched on once and not switched off afterwards. The MFPT of this protocol, $t_0(\gamma,r_0,r_1=0)$, depends on the switch-on rate $r_0$ and is given by (see Section \ref{sec:t0}):
\begin{equation}\label{fgammar0}
\fl t_0(\gamma,r_0,r_1=0)=t_{eq}(\gamma)+\frac{1-e^{-\sqrt{r_0}}}{r_0}-\frac{2e^{-\sqrt{r_0}}\left(\cosh{\sqrt{r_0}}+\frac{\gamma}{\sqrt{r_0}}\sinh{\sqrt{r_0}}-e^{\gamma}\right)}{r_0-\gamma^2}.
\end{equation}
The optimal rate follows from minimizing this expression with respect to $r_0$. We note that the corresponding minimum $t_0^*(\gamma)$ is lower than the Kramers' time $t_{eq}(\gamma)$ for any $\gamma$.

The discontinuous transition originates from the fact that the MFPT $t_0$, unlike $t_1$, is not a unimodal function of the rates but admits two local minima in the $(r_0,r_1)$-plane when $\gamma$ is close to $\gamma_c^{\prime}$. These minima are clear from Figure \ref{fig:t0optgamma}d, which displays $\displaystyle\min_{r_1} t_0(\gamma,r_0,r_1)$ as a function of $r_0$.  For $\gamma\lesssim\gamma_c^{\prime}$, one notices the existence of a metastable minimum at a larger $r_0$, which becomes the absolute minimum when $\gamma\gtrsim\gamma_c^{\prime}$, causing a discontinuity in $r_0^*$. Similarly, the function $\displaystyle\min_{r_0} t_0(\gamma,r_0,r_1)$ admits a metastable minimum at a non-zero $r_1$ when $\gamma\lesssim\gamma_c^{\prime}$.

\subsection{MFPT: optimal potentials at fixed rates}

As our problem has 3 parameters, we can ask another question: Fixing a potential resetting protocol $(r_0, r_1)$, what is the potential strength  that minimizes the MFPT? We define
\begin{equation}
t_1^{\star}(r_0,r_1)=\displaystyle\min_{\gamma} t_1(\gamma,r_0,r_1),
\end{equation}
and $\gamma^\star(r_0,r_1)$ as the corresponding optimum.
\begin{figure}[t]
\centering
\includegraphics[width=\textwidth]{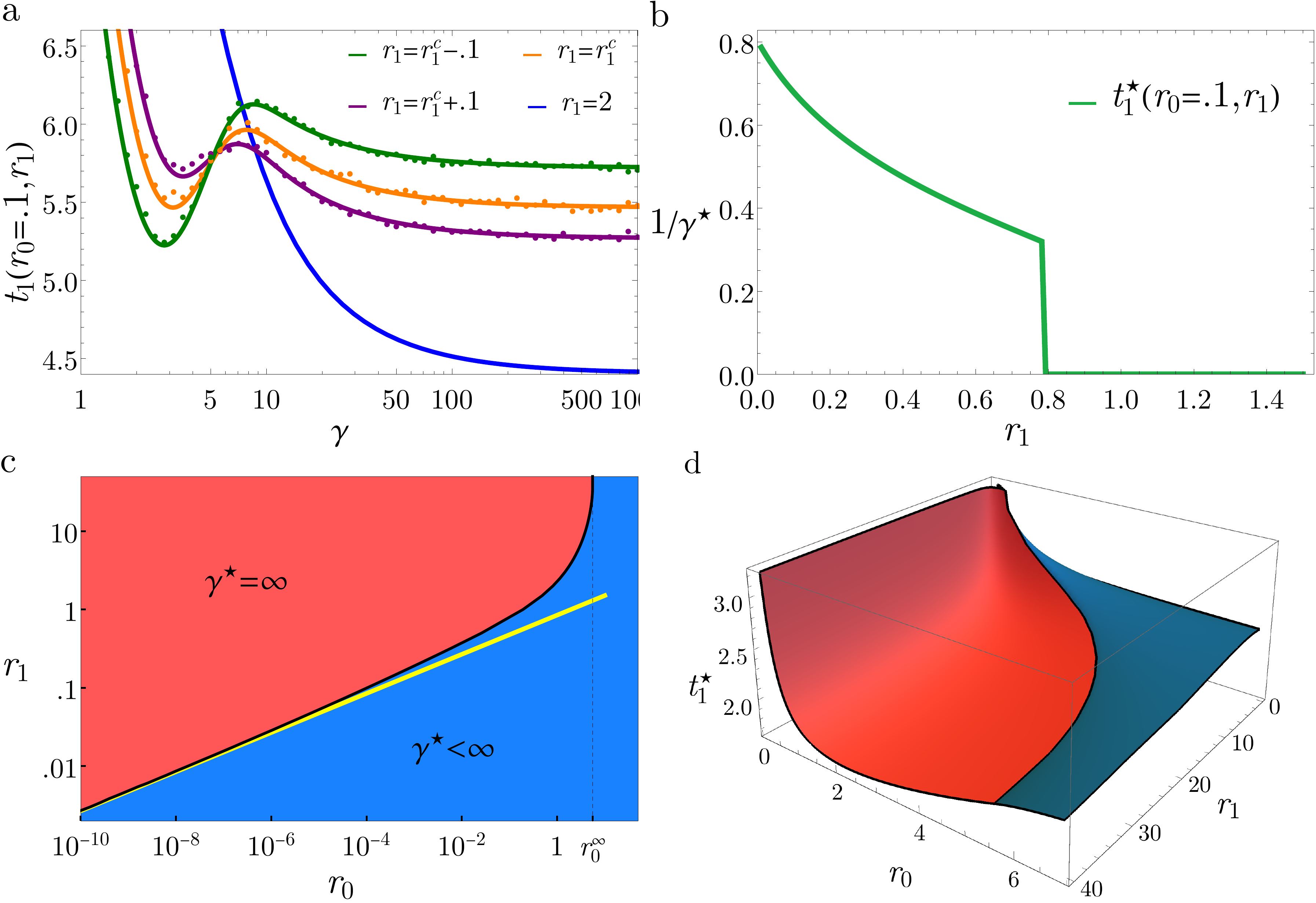}
\caption{
Variations of $t_1$ with the potential strength at fixed rates. (a) Metastability with $\gamma$, at fixed $r_0=0.1$ and $r_1$ around the critical value $r_1^c(r_0=0.1)=0.7854423...$. Symbols represent results from MC simulations; each point averages over $2\times 10^5$ realizations.  (b) Inverse of the optimal strength $\gamma^\star$ vs. $r_1$ deduced from (a). (c) Phase diagram in the $(r_0,r_1)$-plane. The yellow line represents the analytical expression (\ref{critlinesum}) for $r_0\ll 1 $. Above the value $r_0^{\infty}=5.539509...$, for which $r_1^c(r_0^{\infty})=\infty$, the optimal potential strength is always finite.  (d) Optimized MFPT as a function of $(r_0,r_1)$.
}
\label{fig:t1optr0r1}
\end{figure}
Figure \ref{fig:t1optr0r1}a, obtained from the exact solution, shows that $\gamma^\star$ is either finite or infinite. Fixing $r_0$ and a small enough $r_1$, the mean time $t_1(\gamma,r_0,r_1)$ is non-monotonous with $\gamma$ and admits an absolute minimum at a finite value. This minimum becomes metastable and further disappears as $r_1$ increases. In those cases, the MFPT reaches its minimum at $\gamma=\infty$. The agreement with numerical simulations of the Langevin equation (\ref{ruleX1}) is very good. 
The optimal potential strength $\gamma^\star$ (or potential width $1/\gamma^{\star}$) is thus discontinuous at a critical value of $r_1$ which depends on $r_0$ and that we denote as $r_1^c(r_0)$, see Fig. \ref{fig:t1optr0r1}b. Figure \ref{fig:t1optr0r1}c displays the transition line  $r_1^c(r_0)$ which separates the two phases in the $(r_0,r_1)$-plane. Figure \ref{fig:t1optr0r1}d shows the variations of $t_1^{\star}$ in this plane.

The transition can be explained qualitatively from the interplay between two competing effects. Infinitely strong potentials are advantageous as they represent the fastest way to bring the particle back to the origin in those fruitless excursions that explore regions far from the target. However, once at the origin, the particle remains immobile as long as the potential is applied, during a mean time $1/r_1$ which is not dedicated to search. When $r_1$ is small enough, the results above tell us that it is more convenient to choose a finite potential strength, that will induce slower returns toward the origin but allow an exploration of space with the potential on. Surprisingly, the infinite strength phase only exists if the switch-on rate $r_0$ remains below a certain value, $r_0^{\infty}=5.539509...$, for which $r_1^c(r_0^{\infty})=\infty$ (Fig. \ref{fig:t1optr0r1}c). Above that point, $\gamma^{\star}$ is finite and with smooth variations.

Another salient feature of the phase diagram of Fig. \ref{fig:t1optr0r1}c lies in the region of small $r_0$ and $r_1$. In Section \ref{sec:meta}, we calculate the transition line analytically in this regime and find a scaling law:
\begin{equation}
   r_1^c(r_0)\simeq 0.8485...r_0^{1/4}. \label{critlinesum}
\end{equation}
Furthermore, in the vicinity of this line, or more generally for $\sqrt{r_0}\ll r_1\ll 1$ and for any $\gamma$ of order unity ($r_1e^{\gamma}/\gamma^2\ll 1$), one obtains the simple expression
\begin{equation}
    t_1(\gamma,r_0,r_1)\simeq 
    \frac{r_1^2}{r_0}\left(\frac{4e^{\gamma}-\gamma-2}{2\gamma^3}\right).\label{t1smallr0r1}
\end{equation}
At fixed rates, minimizing Eq. (\ref{t1smallr0r1}) with respect to $\gamma$ gives $\gamma_{min}=2.82764...$, which corresponds to a local minimum $1.3887... r_1^2/r_0$ for the MFPT. If $r_1<r_1^c(r_0)$, this local minimum is the absolute one and $\gamma^{\star}\simeq\gamma_{min}$. If $r_1>r_1^c(r_0)$, then $\gamma^{\star}=\infty$.

\section{Stationary density}\label{sec:dens}
We now present the derivations of the results, starting with the particle density. Let us introduce $P_\sigma(X,t)$, the probability that the particle is around $X$ and the potential in state $\sigma=\{0,1\}$ at time $t$, the initial conditions being implicit. The two densities satisfy the forward Fokker-Planck equations (with unit friction),
\begin{eqnarray}
\fl     \frac{\partial}{\partial t}P_0(X,t)=D\frac{\partial^2}{\partial X^2}P_0(X,t)-R_0 P_0(X,t)+R_1P_1(X,t),\label{dpst0}\\
\fl    \frac{\partial }{\partial t}P_1(X,t)=D\frac{\partial^2}{\partial X^2}P_1(X,t)+\frac{\partial}{\partial X}\left( V^{\prime}(X)P_1(X,t)\right)-R_1 P_1(X,t)+R_0P_0(X,t)\label{dpst1}.
\end{eqnarray}
The first term of the r.h.s. of Eq. (\ref{dpst0}) accounts for free diffusion, whereas the second and third terms represent, respectively, the negative probability flux out of the state $\sigma=0$ (at rate $R_0$) and the positive flux into the same state (at rate $R_1$). Similarly for Eq. (\ref{dpst1}), with an advection term caused by the external potential. 
We employ in the following the dimensionless variables $(x,t)$ and the dimensionless parameters defined at the beginning of Section \ref{sec:sum}, where $L$ is an arbitrary length.  The densities of the random variable $x=X/L$ are $p_0(x,t)$ and $p_1(x,t)$. Denoting $p_{\sigma}^+(x,t)\equiv p_{\sigma}(x,t)$ with $x>0$, we have
\begin{eqnarray}
    \frac{\partial p^+_0}{\partial t}=\frac{\partial^2p^+_0}{\partial x^2}-r_0 p^+_0+r_1p^+_1,\label{pst0}\\
    \frac{\partial p^+_1}{\partial t}=\frac{\partial^2p^+_1}{\partial x^2}+\gamma\frac{\partial p^+_1}{\partial x}-r_1 p^+_1+r_0p^+_0\label{pst1}.
\end{eqnarray}
The densities $p_0^-(x,t)$ and $p_1^-(x,t)$ for $x<0$ obey the same equations, except that $\gamma$ is changed to $-\gamma$.

\begin{figure}
    \centering
    \includegraphics[width=.7\linewidth]{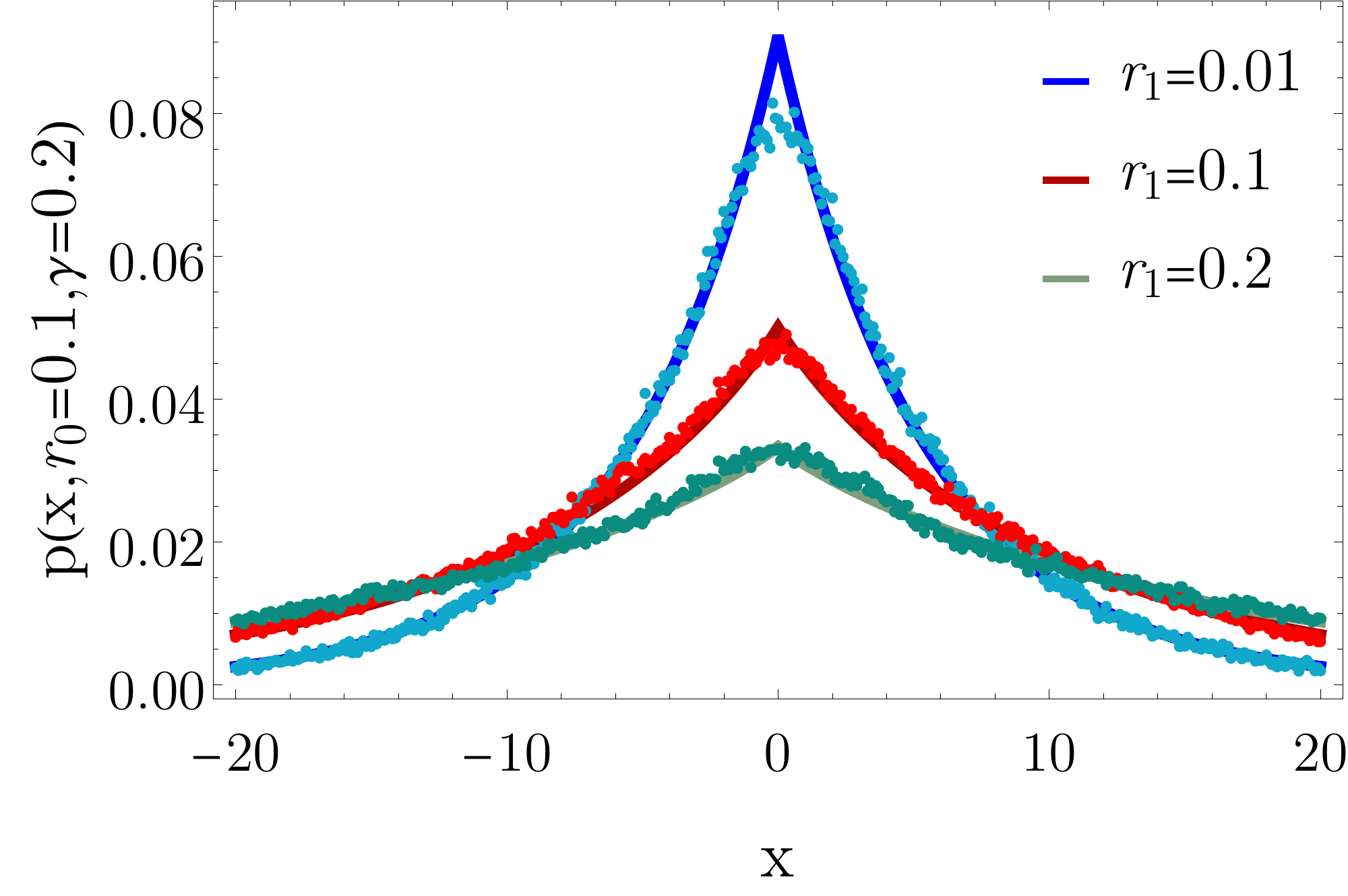}
    \caption{Non-equilibrium steady states. The dots represent Monte Carlo simulations and the solid lines Eq. (\ref{ptotal}).}
    \label{fig:my_label3}
\end{figure}

We consider the infinite unbounded line and explore the existence of non-equilibrium stationary solutions for the joint densities and the total density $p(x)=p_0(x)+p_1(x)$. In the steady state we have $p_{\sigma}(-x)=p_{\sigma}(x)$ by symmetry and it is sufficient to solve for $x>0$. Setting the time derivatives to zero and seeking solutions of Eqs. (\ref{pst0})-(\ref{pst1}) proportional to $\exp(-\lambda x)$, $\lambda$ must satisfy
\begin{equation}\label{char0}
    \left|\begin{array}{cc}
    \lambda^2-r_0 & r_1 \\
    r_0 & \lambda^2-\gamma\lambda-r_1
    \end{array}\right|=0,
\end{equation}
or
\begin{equation}
    \lambda\left[\lambda^3-\lambda^2 \gamma -\lambda(r_1+r_0)+ r_0\gamma\right]=0.\label{charpolySt}
\end{equation}
In addition to the simple root $\lambda_0=0$, there are three other roots \cite{abramowitz}:
\begin{equation}                
    \lambda_k=\frac{1}{3}\left[\gamma+2b\cos\left(\frac{\theta+2(k-1)\pi}{3}\right)\right]\label{lambdas}
\end{equation}
where $k=\{1,2,3\}$ and
\begin{eqnarray}
    b=\sqrt{3(r_0+r_1)+\gamma^2},\label{blambda}\\
    \theta=\arccos\left[\frac{9\gamma(r_1-2r_0)+2\gamma^3}{2(3(r_0+r_1)+\gamma^2)^{3/2}}\right].\label{theta}
\end{eqnarray}
It is relatively straightforward to show that the argument in Eq. (\ref{theta}) is contained in the interval $[-1,1]$ for any non-zero positive $\gamma$, $r_0$ and $r_1$; 
therefore the three roots $\lambda_k$ are real. We next wish to determine their sign and retain only those that are positive to ensure the convergence of the densities as $x\rightarrow+\infty$. The polynomial $P(\lambda)=\lambda^3-\lambda^2 \gamma -\lambda(r_1+r_0)+ r_0\gamma$ is such that $P(0)=r_0\gamma>0$ and $P^{\prime}(0)=-r_1-r_0<0$. Combined to the fact that $P(\lambda)$ decreases over a single finite interval (since it is of degree 3), these inequalities imply that one root must be negative and the other two positive. 
To find the negative root, we notice that $\frac{2(k-1)\pi}{3}\le \frac{\theta+2(k-1)\pi}{3}\le \frac{(2k-1)\pi}{3}$, since $0\le\theta\le\pi$. With $k=2$ the argument of the cosine in Eq. (\ref{lambdas}) is thus in the interval $[\frac{2\pi}{3},\pi]$, which implies that the cosine is smaller than $-1/2$. Since $b>\gamma$ by definition, we conclude that $\lambda_2<0$. In summary,
\begin{equation}
    \lambda_1>0,\ \lambda_3>0,\  \lambda_2<0,\label{eq:uneqs}
\end{equation}
and the acceptable solutions for $p^+_1$ take the form
\begin{equation}
    p^+_1(x)=A_1e^{-\lambda_1 x}+A_3e^{-\lambda_3 x},
\end{equation}
while $p_0^+$ follows from Eq. (\ref{pst1}) with $\partial p_1/\partial t=0$:
\begin{equation}
    p^+_0(x)=\frac{r_1+\lambda_1\left(\gamma-\lambda_1\right)}{r_0}A_1e^{-\lambda_1 x}+\frac{r_1+\lambda_3\left(\gamma-\lambda_3\right)}{r_0}A_3e^{-\lambda_3 x}.
\end{equation}
The constants $A_1$ and $A_3$ are determined from normalization
\begin{equation}
2\int_0^{\infty} dx[p_0^+(x)+p_1^+(x)]=1
\end{equation}
and from the boundary condition
\begin{equation}
    \frac{\partial p^+_1}{\partial x}\Big|_{x=0}=-\gamma p^+_1(x=0).\label{bcst3}
\end{equation}
The latter equality follows from taking the integral $\int_{-\epsilon}^{\epsilon}dx$ of the stationary equation (\ref{dpst1}), using continuity of the densities and noticing that $\partial_x p^+_1(\epsilon)=-\partial_x  p^-_1(-\epsilon)$, $p_1^+(\epsilon)=p_1^-(-\epsilon)$ and $p_0^+(\epsilon)=p_0^-(-\epsilon)$, before taking the limit $\epsilon\rightarrow0$. Consequently, $p_1(x)$ exhibits a cusp at $x=0$. One obtains
\begin{eqnarray}
    A_1=\frac{\lambda _1\lambda _3 \left(\gamma-\lambda_3\right)  r_0}{2 \lambda _2 \left(\lambda _3-\lambda _1\right) \left(r_0+r_1\right)},\label{A1steady}\\
    A_3=-\frac{\lambda _1 \lambda _3 \left(\gamma-\lambda_1\right) r_0}{2 \lambda _2 \left(\lambda _3-\lambda _1\right) \left(r_0+r_1\right)},\label{A3steady}
\end{eqnarray}
where $\lambda_2$ appears as we have used the identity $\lambda_1+\lambda_2+\lambda_3=\gamma$. The total density $p_0(x)+p_1(x)$ reads
\begin{equation}
    p(x)=\frac{r_0\gamma}{2(r_0+r_1)}\left(\frac{\lambda_3(\gamma-\lambda_3)}{\lambda_2(\lambda_3-\lambda_1)}e^{-\lambda_1|x|}-\frac{\lambda_1(\gamma-\lambda_1)}{\lambda_2(\lambda_3-\lambda_1)}e^{-\lambda_3|x|}\right)\label{ptotal}
\end{equation}
for $x\in\mathbb{R}$. It is displayed in Figure \ref{fig:my_label3} in a few examples and compared with numerical simulations of the Langevin equation (\ref{ruleX1}), showing a very good agreement.

One can analyze the behavior of the distribution in the limit of large $\gamma$, where $\lambda_1\simeq\gamma+\frac{r_1}{\gamma}$, $\lambda_2\simeq-\sqrt{r_0}-\frac{r_1}{2\gamma}$ and $\lambda_3\simeq\sqrt{r_0}-\frac{r_1}{2\gamma}$. Inserting these expressions into Eq. (\ref{ptotal}) gives
\begin{equation}
     p(x)\simeq \frac{ r_0\gamma}{2(r_0+r_1)} e^{-\gamma |x|}+\frac{ \sqrt{r_0}r_1}{2(r_0+r_1)} e^{-\sqrt{r_0} |x|},\label{eq:denslargegamma2}
\end{equation}
which is expression (\ref{eq:denslargegamma}).
In the limit $\gamma=\infty$, this density becomes
\begin{equation}
      p_{\gamma=\infty}(x)=\frac{r_0}{r_0+r_1}\delta(x)+ \frac{ \sqrt{r_0}r_1}{2(r_0+r_1)} e^{-\sqrt{r_0} |x|}.
\end{equation}
We recover the NESS of a resetting process with refractory periods, {\it i.e.}, in which each resetting is instantaneous and followed by a period of immobility at the origin, of duration exponentially distributed \cite{evans2018effects}.
If we then take the limit $r_1\rightarrow\infty$, corresponding to a fast potential switched-off after attracting the particle to the origin, the refractory period disappears and
\begin{equation}
      p_{\gamma=\infty,r_1=\infty}(x)= \frac{ \sqrt{r_0}}{2} e^{-\sqrt{r_0} |x|},
\end{equation}
which coincides with the NESS of a particle with unit diffusion constant and resetting rate $r_0$ to the origin \cite{evans2011diffusion}.

\section{Mean first passage times}\label{sec:mfpt}

In dimensionless units, an absorbing boundary is now placed at $-1$ and we define $Q_{0}(x,t)$ as the probability that the particle has not hit the boundary up to time $t$, given an initial position $x$ and initial potential state $\sigma(t=0)=0$. Similarly, $Q_{1}(x,t)$ corresponds to a potential initially on. These survival probabilities satisfy the backward Fokker-Planck equations \cite{risken1984fokker}
 \begin{eqnarray}
     \frac{\partial Q_0}{\partial t}=\frac{\partial^2 Q_0}{\partial x^2}+r_0(Q_1-Q_0),\label{FPQP1}\\
      \frac{\partial Q_1}{\partial t}=\frac{\partial^2 Q_1}{\partial x^2}- v^{\prime}(x)\frac{\partial Q_1}{\partial x} +r_1(Q_0-Q_1),\label{FPQP2}
\end{eqnarray}
where $v(x)=\gamma|x|$ is the dimensionless potential. The derivation of similar coupled backward equations can be found in \cite{benichou2011intermittent,malakar2018steady,mercado2019first,bressloff2020switching} for other switching processes or in the context of intermittent search.
Let us define $Q^+(x,t)\equiv Q(x,t)$ with $x>0$, $Q^-(x,t)\equiv Q(x,t)$ with $-1\le x<0$, and introduce the Laplace transform $\widetilde{f}(x,s)=\int_0^{\infty}dt f(x,t)e^{-st}$. In the Laplace domain, Eqs. (\ref{FPQP1})-(\ref{FPQP2}) read, for $x>0$,
\begin{eqnarray}
     \frac{\partial^2 \widetilde{Q}^+_0}{\partial x^2}+r_0\widetilde{Q}^+_1-(r_0+s)\widetilde{Q}^+_0=-1,\label{laplacesystem1}\\
      \frac{\partial^2 \widetilde{Q}^+_1}{\partial x^2}-\gamma\frac{\partial \widetilde{Q}^+_1}{\partial x} +r_1\widetilde{Q}^+_0-(r_1+s)\widetilde{Q}^+_1=-1,\label{laplacesystem2}
\end{eqnarray}
where we have used the initial condition $Q^+_{0,1}(x,t=0)=1$.  We notice at this point that a system like  (\ref{laplacesystem1})-(\ref{laplacesystem2}) is difficult to solve exactly. Whereas the inhomogeneous solution is given by $\widetilde{Q}_0^+=\widetilde{Q}_1^+=1/s$, looking for homogeneous solutions of the form $e^{\lambda x}$ yields the equation
\begin{equation}\label{char1}
    \left|\begin{array}{cc}
    \lambda^2-(r_0+s) & r_0 \\
    r_1 & \lambda^2-\gamma\lambda-(r_1+s)
    \end{array}\right|=0,
\end{equation}
which requires to find the (negative) roots $\lambda(s)$ of a 4th order polynomial. 

An important simplification occurs in the case $s=0$, however, as the degree of the polynomial reduces to 3 and Eq. (\ref{char1}) becomes identical to Eq. (\ref{charpolySt}), which has been already solved for the densities $p_0^+(x)$ and $p_1^+(x)$. In dimensionless units, 
$\widetilde{Q}_0^+(x,s=0)$ is the mean first passage time rescaled by $L^2/D$, denoted as $t_0^+(x)$. We deduce from Eqs. (\ref{laplacesystem1})-(\ref{laplacesystem2}) the equations for the rescaled MFPTs $t_0(x)$ and $t_1(x)$ for $x>0$:
\begin{eqnarray}
     \frac{\partial^2 t^+_0(x)}{\partial x^2}+r_0[t^+_1(x)-t^+_0(x)]=-1,\label{Tsystem1}\\
      \frac{\partial^2 t^+_1(x)}{\partial x^2}-\gamma\frac{\partial t^+_1(x)}{\partial x} +r_1[t^+_0(x)-t^+_1(x)]=-1,\label{Tsystem2}
\end{eqnarray}
whereas for $-1\le x<0$, we have
\begin{eqnarray}
     \frac{\partial^2 t^-_0(x)}{\partial x^2}+r_0[t^-_1(x)-t^-_0(x)]=-1,\label{Tsystem1minus}\\
      \frac{\partial^2 t^-_1(x)}{\partial x^2}+\gamma\frac{\partial t^-_1(x)}{\partial x} +r_1[t^-_0(x)-t^-_1(x)]=-1.\label{Tsystem2minus}
\end{eqnarray}
In \ref{sec:exact}, we solve these equations on each side and match them at $x=0$, through the continuity of the MFPTs and their derivatives. There are six boundary conditions:
\begin{eqnarray}
    t^+_\sigma(x=0)=t^-_\sigma(x=0),\label{BC1}\\
    \partial_x t^+_\sigma\big|_{x=0}=
    \partial_x t^-_\sigma\big|_{x=0},\label{BC2}\\
     t^-_\sigma(x=-1)=0,\label{BC3}
\end{eqnarray}
where $\sigma=\{0,1\}$ and the last condition enforces absorption at $x=-1$. The solutions for $t_0(x=0)$ and $t_1(x=0)$ are given explicitly by Eqs. (\ref{solt1})-(\ref{A0p}). We may also take the average over the initial potential state to obtain an averaged MFPT:
\begin{equation}\label{tav}
t_{av}=\frac{r_0}{r_0+r_1} t_1 +\frac{r_1}{r_0+r_1} t_0.
\end{equation}

\subsection{Minimization with respect to $r_0$ and $r_1$}

\begin{figure}[t]
    \centering
     \includegraphics[width=\textwidth]{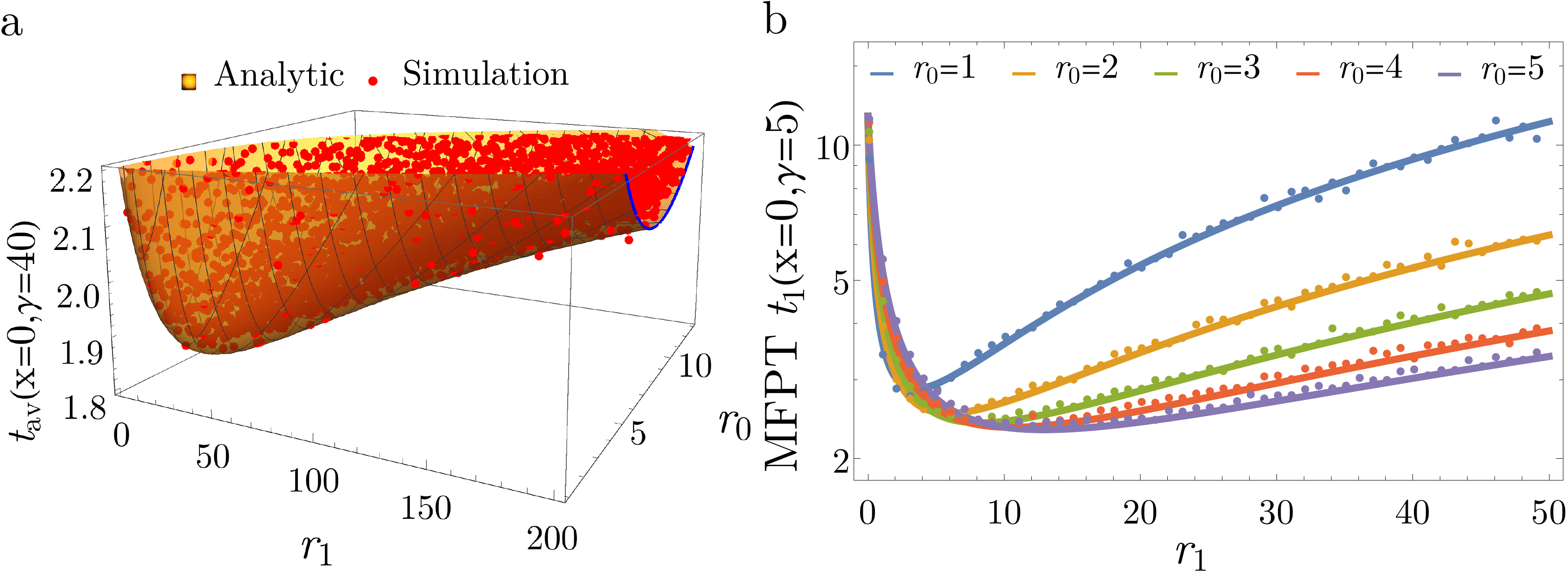}
    \caption{a) Averaged MFTP for a particle starting at the origin as a function of ($r_0,r_1$), with $\gamma=40$. The red points represent simulation results and the yellow surface the exact solution. b) MFPT $t_1$ as a function of $r_1$ for several values of $r_0$ and a fixed $\gamma=5$. The points are simulation results and the solid line the exact solution.}
     \label{fig:tminiplane}
\end{figure}

In this part, we focus on the behavior of the solutions in the $(r_0,r_1)$-plane, at fixed potential strength $\gamma$. The quantity $t_{av}(x=0)$ exhibits a global minimum $(r_1^*,r_0^*)$, as illustrated in the example of Figure  \ref{fig:tminiplane}a with $\gamma=40$. The surface is obtained from numerical evaluations of the exact solution and the agreement with Monte Carlo simulations is very good. Similar results are observed for $t_0(x=0)$ and $t_1(x=0)$ separately (not shown). Figure \ref{fig:tminiplane}b displays the MFPT $t_1(x=0)$ as a function of $r_1$ at fixed $r_0$. A minimum is reached at a finite $r_1$, and the curves with different $r_0$ reach the same value at $r_1=0$, as they should: this point corresponds to the equilibrium dynamics where the potential is present from $t=0$ and where $r_0$ plays no role. 


\subsubsection{Naive argument:}\label{sec:naive}

A heuristic argument can explain qualitatively why a second order transition for the optimal parameters of $t_1(x=0)$ occurs as $\gamma$ is varied. It consists in replacing the problem by that of a particle in a steady environment given by the mean potential felt by the original particle, {\it i.e.}, $\bar{v}(x)=\bar{\gamma}|x|$ with $\bar{\gamma}=r_0\gamma/(r_0+r_1)\le\gamma$. In this effective description, the MFPT satisfies the equation
\begin{equation}\label{heurt}
    \frac{\partial^2 \bar{t}}{\partial x^2}-\frac{\partial \bar{v}(x)}{\partial x}\frac{\partial \bar{t}}{\partial x}=-1,
\end{equation}
The solution of Eq. (\ref{heurt}) evaluated at $x=0$ is given by Eq. (\ref{t1eq}), where $\gamma$ has to be replaced by $\bar{\gamma}$:
\begin{equation}\label{soltheur}
    \bar{t}(x=0)=t_{eq}(\bar{\gamma}).
\end{equation}
As mentioned earlier, the MFPT $t_{eq}$ is a non-monotonous function of its argument and reaches a minimum at $\gamma_{eq}^*=1.244678...$. Hence, at fixed $\gamma$ there are two ways of minimizing Eq. (\ref{soltheur}): If $\gamma<\gamma_{eq}^*$, on the decreasing side of the curve, the argument $\bar{\gamma}$ should be as large as possible, {\it i.e.}, $\bar{\gamma}=\gamma$, which implies $r_1^*=0$. But if $\gamma>\gamma_{eq}^*$, $\bar{\gamma}$ can be tuned to exactly match the optimal parameter $\gamma_{eq}^*$ by choosing the rates such that:
\begin{equation}
    \gamma_{eq}^*=\frac{r_0^*(\gamma)}{r_0^*(\gamma)+r_1^*(\gamma)}\gamma.
\end{equation}
The above relation implies $r_1^*(\gamma)\neq0$, since $\gamma_{eq}^*<\gamma$. Hence, a transition between optimal protocols with zero and non-zero switch-off rates would occur at $\gamma=\gamma_{eq}^*$. It is reasonable to think that when the potential is very confining, one needs to switch it off from time to time to let the particle find the target.

However, the above description is incorrect, as the true critical strength $\gamma_c$ is lower than $\gamma_{eq}^*$, albeit by less than two percents, a feature suggestive of a more complex mechanism. In addition, the above reasoning does not explain why the optimal MFPT $t_1^*(\gamma)$ is in general below $t_{eq}(\gamma_{eq}^*)$ after the transition.

\subsubsection{Exact perturbation theory for $t_1$:}\label{sec:perturb}

A comprehensive description of the second order transition can be obtained from a perturbative calculation assuming $r_1\ll r_0$. This method turns out to be simpler and physically more intuitive than expanding the exact solution itself, whose long expression is rather tedious to handle (but obviously useful for numerical evaluations and checks).

\vspace{0.5cm}
{\it a) Calculation of $\gamma_c$ and $r_0^*(\gamma_c)$}:
\vspace{0.2cm}

Combining Eqs. (\ref{Tsystem1})-(\ref{Tsystem2minus}), one obtains two fourth-order differential equations for $t^+_1(x)$ and $t^-_1(x)$:\footnote{It is easy to see that the effective description of Eq. (\ref{heurt}) becomes exact when $r_0$ and $r_1$ tend to $\infty$,  $r_1/r_0$ being fixed.}
\begin{eqnarray}
    \frac{\partial^4 t^+_1(x)}{\partial x^4}-\gamma\frac{\partial^3 t^+_1(x)}{\partial x^3}-(r_1+r_0)\frac{\partial^2 t^+_1(x)}{\partial x^2}+r_0\gamma\frac{\partial t^+_1(x)}{\partial x}=r_1+r_0,\label{difT12}\\
    \frac{\partial^4 t^-_1(x)}{\partial x^4}+\gamma\frac{\partial^3 t^-_1(x)}{\partial x^3}-(r_1+r_0)\frac{\partial^2 t^-_1(x)}{\partial x^2}-r_0\gamma\frac{\partial t^-_1(x)}{\partial x}=r_1+r_0\label{difT1m2}.
\end{eqnarray}
Defining the operator $\mathscr{L}_\gamma\equiv \frac{\partial^2}{\partial x^2}-\gamma\frac{\partial}{\partial x}$ and introducing the parameter $\epsilon\equiv\frac{r_1}{r_0}$, Eqs. (\ref{difT12})-(\ref{difT1m2}) become
%
\begin{eqnarray}
\left[\frac{1}{r_0}\frac{\partial^2}{\partial x^2}-1\right]\mathscr{L}_\gamma t^{+}_1-\epsilon\frac{\partial^2 t^{+}_1}{\partial x^2}=1+\epsilon,\label{operatort}\\
\left[\frac{1}{r_0}\frac{\partial^2}{\partial x^2}-1\right]\mathscr{L}_{-\gamma} t^{-}_1-\epsilon\frac{\partial^2 t^{-}_1}{\partial x^2}=1+\epsilon.\label{operatortm}
\end{eqnarray}
To solve these PDEs we write the general solution as an expansion in powers of $\epsilon$ near $\epsilon=0$, or $t_1(x)=t_1^{(0)}(x)+\epsilon t_1^{(1)}(x)+\epsilon^2 t_1^{(2)}(x)+\dots$, where $t_1^{(0)},t_1^{(1)},...$ are functions to be determined on each side. We denote $t_1^{(0)}(x,+)=t_1^{(0)}(x)$ for $x>0$, and  $t_1^{(0)}(x,-)=t_1^{(0)}(x)$ for $-1\le x<0$ [and similarly for the higher orders $t_1^{(1)}(x),t_1^{(2)}(x)$,...].

At leading order:
\begin{eqnarray}
    \left[\frac{1}{r_0}\frac{\partial^2}{\partial x^2}-1\right]\mathscr{L}_{\gamma} t^{(0)}_1(x,+)=1,\label{order0}\\
     \left[\frac{1}{r_0}\frac{\partial^2}{\partial x^2}-1\right]\mathscr{L}_{-\gamma} t^{(0)}_1(x,-)=1.\label{order0m}
\end{eqnarray}
These relations are satisfied if $\mathscr{L}_{\gamma} t^{(0)}_1(x,+)=-1$ and  $\mathscr{L}_{-\gamma} t^{(0)}_1(x,-)=-1$, which are the equations for the mean time $t_{eq}(x)$ in the steady potential $\gamma|x|$. Its expression is obtained by applying the boundary conditions (\ref{BC1})-(\ref{BC3}) (see, {\it e.g.}, \cite{evans2013optimal}):
\begin{eqnarray}
     t_1^{(0)}(x,+)=\frac{2(e^{\gamma}-1)-(1-x)\gamma}{\gamma^2},\label{t10p}\\
     t_1^{(0)}(x,-)=\frac{2(e^{\gamma}-e^{-\gamma x})-(x+1)\gamma}{\gamma^2}.\label{t10m}
 \end{eqnarray}
%
%
%
A relation allowing the exact determination of the critical point $\gamma_c$ can be obtained at the next order in $\epsilon$. From Eqs. (\ref{operatort})-(\ref{operatortm}),
\begin{eqnarray}
\left[\frac{1}{r_0}\frac{\partial^2}{\partial x^2}-1\right]\mathscr{L}_\gamma t^{(1)}_1(x,+)=1+\frac{\partial^2 t^{(0)}_1(x,+)}{\partial x^2},\label{eqt11}\\
\left[\frac{1}{r_0}\frac{\partial^2}{\partial x^2}-1\right]\mathscr{L}_{-\gamma} t^{(1)}_1(x,-)=1+\frac{\partial^2 t^{(0)}_1(x,-)}{\partial x^2}.\label{eqt11m}
\end{eqnarray}

Solving for $t_1^{(1)}(x,+)$, the homogeneous part of (\ref{eqt11}) is given by a constant plus a linear combination of $e^{-\sqrt{r_0}x}$, $e^{\sqrt{r_0}x}$ and $e^{\gamma x}$, where the coefficients of the last two terms must be $0$ to avoid exponential divergence at $x=+\infty$. From Eq. (\ref{t10p}), the right-hand-side of (\ref{eqt11}) is $1$, which yields $x/\gamma$ for the inhomogeneous solution. Similarly, the homogeneous part of $t_1^{(1)}(x,-)$ is a constant plus a linear combination of $e^{-\sqrt{r_0}x}$, $e^{\sqrt{r_0}x}$ and $e^{-\gamma x}$, which are all acceptable ($-1\le x<0$). The right-hand-side of (\ref{eqt11m}) is obtained from Eq. (\ref{t10m}) as $1-2e^{-\gamma x}$, which yields the inhomogeneous solution $-[2r_0xe^{-\gamma x}/(r_0-\gamma^2)+x]/\gamma$. To sum up, $t_1^{(1)}(x)$ takes the form
\begin{eqnarray}
     t_1^{(1)}(x,+)=C^+ + A^+e^{-\sqrt{r_0}x}+\frac{x}{\gamma},\label{t11p} \\
     t_1^{(1)}(x,-)=C^-+ A^-e^{-\sqrt{r_0}x}+B^-e^{\sqrt{r_0}x}+D^- e^{-\gamma x}-\frac{2r_0xe^{-\gamma x}}{\gamma\left(r_0-\gamma^2\right)}-\frac{x}{\gamma}. \label{t11m}
 \end{eqnarray}
Three relations between the six unknowns above are given by the boundary conditions: $t_1^{(1)}(0,+)=t_1^{(1)}(0,-)$, $\partial_x t_1^{(1)}|_{0,+}=\partial_x t_1^{(1)}|_{0,-}$ and $t_1^{(1)}(-1,-)=0$. The remaining three relations stem from the same conditions applied to $t_0^{(0)}(x)$ in the $\epsilon$-expansion of $t_0(x)=t_0^{(0)}+\epsilon t_0^{(1)}...$. Indeed, once $t_1$ is known, $t_0$ is completely determined on each side through the general relations Eqs. (\ref{Tsystem2}) and (\ref{Tsystem2minus}), or
\begin{eqnarray}
    t_0^+(x)=t_1^+(x)-\frac{1}{r_1}\left[\mathscr{L}_{\gamma}t_1^+(x)+1\right],\label{eq:relt0t1}\\
    t_0^-(x)=t_1^-(x)-\frac{1}{r_1}\left[\mathscr{L}_{-\gamma}t_1^-(x)+1\right]
    .\label{eq:relt0t1m}
\end{eqnarray}
At order $1/\epsilon$, one recovers: $\mathscr{L}_{\pm\gamma} t^{(0)}_1(x,\pm)=-1$, which was already solved. At order $\epsilon^{0}$,
\begin{eqnarray}
    t_0^{(0)}(x,+)=t^{(0)}_1(x,+)-\frac{1}{r_0}\mathscr{L}_{\gamma}t^{(1)}_1(x,+),\label{t00p}\\
    t_0^{(0)}(x,-)=t^{(0)}_1(x,-)-\frac{1}{r_0}\mathscr{L}_{-\gamma}t^{(1)}_1(x,-).\label{t00m}
\end{eqnarray}
When applying the boundary conditions to $t_0^{(0)}$ given by Eqs. (\ref{t00p})-(\ref{t00m}), we obtain 3 new relations that only involve $A^+$, $A^-$ and $B^-$, which are readily solved and substituted into the other $3$ relations for the remaining coefficients. After some algebra, one obtains, after evaluating at $x=0$: 
\begin{equation}\label{t1approx1}
\fl \eqalign{
    t^{(1)}_1(\gamma,r_0)=&-\gamma\frac{d t_{eq}(\gamma)}{d \gamma}+\frac{\sqrt{r_0}e^{- \sqrt{r_0}}\left[2\left(e^{\gamma }-1\right)\left(e^{- \sqrt{r_0}}+\frac{\gamma}{\sqrt{r_0}}\right)+4 e^{\gamma }-1+\frac{\gamma ^2}{r_0}\right]}{\left(\gamma +\sqrt{r_0}\right) \left(r_0-\gamma
   ^2\right)}\\
   &+\frac{2 \left(e^{\gamma }-1\right)-2 \gamma e^{\gamma }-\frac{\gamma
   }{\sqrt{r_0}}+e^{-2 \sqrt{r_0}}}{r_0-\gamma ^2}+\frac{4 r_0e^{\gamma }\left(\frac{\gamma}{\sqrt{r_0}}- e^{\gamma-\sqrt{r_0}}\right)}{\left(r_0-\gamma ^2\right)^2},}
\end{equation}
whereas the complete expressions of $t_1^{(1)}(x)$ for $x>0$ and $-1\le x<0$ are written in \ref{appendixt2}.

Although the expression (\ref{t1approx1}) looks intricate, the determination of the critical potential strength $\gamma_c$ is quite simple. At fixed $\gamma$, if $t_1^{(1)}$ is positive for all $r_0$, then applying a resetting protocol with a small $r_1$ will always cause an increase of $t_1$, as the first correction will be positive (recall that $\epsilon\ge0$). On the other hand, if at fixed $\gamma$ there exists a range of $r_0$ such that $t_1^{(1)}<0$, then potential resetting can expedite target encounter. The critical value $\gamma_c$ is given by the marginal situation between these two cases. Figure \ref{fig:t1optgamma}d shows $t_1^{(1)}$ as a function of $r_0$ for a few values of $\gamma$. This quantity plays the role of a \lq\lq dispersion relation\rq\rq, in analogy with instabilities in pattern formation problems: it is non-monotonic but keeps the same sign for $\gamma$ below a threshold, whereas for
\begin{equation}
\gamma=\gamma_c\equiv1.228780...,
\end{equation}
the curve becomes tangent to the horizontal axis $y=0$ at some point $r_c$. This marginal rate is given by
\begin{equation}
r_c\equiv41.969027..., 
\end{equation}
which is the value taken by the optimal rate $r_0^*$ at the transition, as announced in Eq. (\ref{r0opt1gammac}). Setting $\gamma-\gamma_c>0$ but small, $t_1^{(1)}(\gamma,r_0)$ will present negative values over a small interval centered around $r_c$, hence one must choose $r_0\simeq r_c$ to minimize the MFPT at $\epsilon$ fixed. A similar reasoning applies to the fastest growing mode in pattern formation problems.

\vspace{0.5cm}
{\it b) Behaviour of $t^*_1$ and $r_1^*$ near $\gamma_c$}:
\vspace{0.2cm}

Setting $x=0$, we have shown that for $\gamma\gtrsim\gamma_c$ and $r_0\simeq r_c$, the series expansion of the MFPT in powers of $\epsilon=r_1/r_0$,
\begin{equation}\label{sumexp1}
t_1(\gamma,\epsilon,r_0)=t_{eq}(\gamma)+\epsilon t_1^{(1)}(\gamma,r_0)+\epsilon^2 t_1^{(2)}(\gamma,r_0)+\dots,
\end{equation}
is such that $t_1^{(1)}<0$. As $|t_1^{(1)}|$ is small near the transition, if the next order term $t_1^{(2)}$ is positive, one can carry out the minimization of the MFPT with respect to $\epsilon$, while the terms of order $\epsilon^3$ and higher can be safely neglected. In the spirit of the Ginzburg-Landau theory of phase transitions, we expand the different coefficients in Eq. (\ref{sumexp1}) near ($\gamma_c,r_c$).  As the critical point fulfills $t_1^{(1)}(\gamma_c,r_c)=0$ and $\partial_{r_0}t_1^{(1)}|_{\gamma_c,r_c}=0$, Eq. (\ref{sumexp1}) becomes
\begin{equation}\label{GL}
 \fl   t_1(\gamma,\epsilon,r_0)=t_{eq}(\gamma_c)+(\gamma-\gamma_c)\left.\frac{\partial t_{eq}}{\partial \gamma}\right|_{\gamma_c}+
    \epsilon 
    (\gamma-\gamma_c)\left.\frac{\partial t_{1}^{(1)}}{\partial \gamma}\right|_{\gamma_c,r_c}+\epsilon^2t^{(2)}_1(\gamma_c,r_c)+...
\end{equation}
at order $\gamma-\gamma_c$. Minimization of the last two terms at fixed $\gamma$ gives $\epsilon^*$, or equivalently:
\begin{equation}
    r_1^*(\gamma)=-(\gamma-\gamma_c)r_c\left.\frac{\partial t_1^{(1)}}{\partial \gamma}\right|_{\gamma_c,r_c}[2t_1^{(2)}(\gamma_c,r_c)]^{-1}+o\left(\gamma-\gamma_c\right)\label{optimalr12},
\end{equation}
where we have replaced $r_0$ by $r_c$ at leading order in $\gamma-\gamma_c$.  With the help of Mathematica$^{\circledR}$, the calculation of $t_1^{(2)}$ can be performed similarly to that of $t_{1}^{(1)}(x)$ at order $\epsilon$, and its expression for $x=0$ is given in \ref{appendixt2}. Importantly, we find out that $t_1^{(2)}(\gamma_c,r_c)$ is positive, as needed, and the numerical evaluation of (\ref{optimalr12}) gives,
\begin{equation}
    r_1^*(\gamma)= 32.91301557...(\gamma-\gamma_c)+o(\gamma-\gamma_c)\label{optimalr12bis}
\end{equation}
which is Eq. (\ref{r1opt1gammac}). Therefore, the dimensionless order parameter $r_1^*$ grows linearly near $\gamma_c$ with a large prefactor. Setting $\epsilon^*=c(\gamma-\gamma_c)$ 
and substituting into  Eq. (\ref{GL}), we obtain the minimal MFPT
\begin{equation}\label{eq2ndt1}
\fl    t^*_1(\gamma)=t_{eq}(\gamma_c)+(\gamma-\gamma_c)\left.\frac{\partial t_{eq}}{\partial \gamma}\right|_{\gamma_c}+(\gamma-\gamma_c)^2\left\{\frac{1}{2}\left.\frac{\partial^2 t_{eq}}{\partial \gamma^2}\right|_{\gamma_c}+c\left.\frac{\partial t_1^{(1)}}{\partial \gamma}\right|_{\gamma_c,r_c}+c^2t^{(2)}_1(\gamma_c,r_c)\right\}+...
\end{equation}
where $t_{eq}$ has been expanded further. It is clear from (\ref{eq2ndt1}) that the first derivative of $t_1^*(\gamma)$ is continuous across the transition, whereas the second derivative presents a discontinuity. Numerical evaluation of the prefactors gives
\begin{equation}\label{t1optcritbis}
 \fl    t^*_1(\gamma)= t_{eq}(\gamma_c)-0.0225432...\left(\gamma-\gamma_c\right)-0.00103404...\left(\gamma-\gamma_c\right)^2+o\left((\gamma-\gamma_c)^2\right),
\end{equation}
which is Eq. (\ref{t1optcrit}). We have checked that this expression is in very good agreement with the exact solution near $\gamma_c$. 

\subsubsection{Calculation of $t_0(\gamma,r_0,r_1=0)$:}\label{sec:t0}

We can further insert the solutions (\ref{t10p})-(\ref{t10m}) for $t_1^{(0)}(x,\pm)$ and (\ref{t11p})-(\ref{t11m}) for
$t_1^{(1)}(x,\pm)$ into Eq. (\ref{t00p})-(\ref{t00m}), to obtain the exact expression of $t_0^{(0)}$, the MFPT in the absence of potential at $t=0$ and with $r_1=0$. In this case, the potential is applied once at rate $r_0$ and is never switched off afterwards. One has:
\begin{eqnarray}
\fl \eqalign{
    t^{+}_0(\gamma,r_0,r_1=0,x)=&\frac{2(e^{\gamma}-1)-(1-x)\gamma}{\gamma^2}+\frac{1-e^{-\sqrt{r_0}(x+1)}}{r_0}\\
    &-\frac{2e^{-\sqrt{r_0}(x+1)}
   \left(\frac{\gamma}{\sqrt{r_0}}\sinh{\sqrt{r_0}}+\cosh{\sqrt{r_0}}-e^{\gamma}\right)}{r_0-\gamma^2},}\\
\fl \eqalign{ t^{-}_0(\gamma,r_0,r_1=0,x)=&\frac{2\left( e^{\gamma}-e^{-x \gamma}\right)-(1+x)\gamma}{\gamma^2}+\frac{1-e^{-\sqrt{r_0} (x+1)}}{r_0}\\
   &-\frac{2e^{-\sqrt{r_0}}\left(\left(\frac{\gamma}{\sqrt{r_0}}-1\right)\sinh{\left[\sqrt{r_0}(x+1)\right]}+e^{-x \gamma+\sqrt{r_0}}-e^{\gamma-\sqrt{r_0}x}\right)}{r_0-\gamma^2}.}
\end{eqnarray}
At $x=0$, the above relations read
\begin{equation}
 \fl   t_0(\gamma,r_0)=t_{eq}(\gamma)+\frac{1-e^{-\sqrt{r_0}}}{r_0}-\frac{2e^{-\sqrt{r_0}}
   \left(\frac{\gamma}{\sqrt{r_0}}\sinh{\sqrt{r_0}}+\cosh{\sqrt{r_0}}-e^{\gamma}\right)}{r_0-\gamma^2},
\end{equation}
as announced in Eq. (\ref{fgammar0}).
At fixed $\gamma$, this expression can be minimized with respect to $r_0$, yielding the optimal quantities $r_0^*$ and $t_0^*$ for $\gamma<\gamma_c^\prime$, which are displayed in Figs. \ref{fig:t0optgamma}a and \ref{fig:t0optgamma}c. In its present form, the perturbative theory does not allow to study the first order transition in $t_0^*$, though. So far we have not found a simple way to calculate $\gamma_c^{\prime}$.

\subsection{Minimization with respect to the potential strength}\label{sec:meta}

In this Section, 
we perform a minimization of the MFPT $t_1$ with respect to the potential strength, holding the rates of the intermittent dynamics fixed.
The phase diagram depicted in Figure \ref{fig:t1optr0r1}c is obtained from numerical minimization of the exact expression for $t_1(\gamma,r_0,r_1,x=0)$ derived in \ref{sec:exact}. Here, we wish to better understand the phenomenon of metastability and the shape of the transition line in the $(r_0,r_1)$-plane, at which the optimal strength $\gamma^\star$ changes from finite to infinite values. Noticing that $r_1$ is often much larger than $r_0$ at the transition, especially in the small rate region, we use this property to obtain an approximate expression of the MFPT.


We start by deducing below the MFPT with $\gamma=\infty$ (at any rates) from a backward Fokker-Planck equation, and check the agreement with the full exact expression in the limit $\gamma\to\infty$. In a second step, we take $\gamma$ finite (of order 1) and obtain from the full solution a simplified expression for $t_1$ within the assumption $r_0\ll r_1\ll 1$, showing explicitly the existence of a local minimum at a finite potential strength. Finally, these two mean times are compared to obtain the absolute minimum and the transition line in the small $r_0$ regime. 

If the potential has infinite strength, the diffusive particle returns to the origin infinitely fast once the potential is turned on, and remains still during a random time of mean $1/r_1$ until the next restart. Hence, the MFPT with the initial condition $\sigma(t=0)=1$ is independent of $x$, and longer by an amount $1/r_1$ than the MFPT with $\sigma(t=0)=0$ and $x=0$: 
\begin{equation}
    t_1(\gamma=\infty,x)=t_0(\gamma=\infty,x=0)+\frac{1}{r_1}.\label{t1x0gaminf}
\end{equation}
Substituting this expression into Eq. (\ref{Tsystem1}) one obtains 
\begin{equation}
 \fl   \frac{\partial^2 t^{+}_0(\gamma=\infty,x)}{\partial x^2}-r_0 t^{+}_0(\gamma=\infty,x)=-\frac{r_0+r_1}{r_1}-r_0t^{+}_0(\gamma=\infty,x=0),\label{BFPt0gaminf}
\end{equation}
as well as a similar equation for $t^{-}_0(\gamma=\infty,x)$.
In Eq. (\ref{BFPt0gaminf}), we notice that the function $\frac{r_1}{r_0+r_1}t_0^{+}(\gamma=\infty,x)$ satisfies the same backward equation than the MFPT in diffusion with resetting at rate $r_0$ without refractory period, whose expression reads $(e^{\sqrt{r_0}-1})/r_0$ in our dimensionless units and for $x=0$ \cite{evans2011diffusion}. We deduce
\begin{equation}
    t_0(\gamma=\infty,x=0)=\frac{r_0+r_1}{r_0 r_1}\left(e^{\sqrt{r_0}}-1\right),
\end{equation}
and, from Eq. (\ref{t1x0gaminf}),
  \begin{equation}
     t_1(\gamma=\infty)=\frac{e^{\sqrt{r_0}}}{r_1}+\frac{e^{\sqrt{r_0}}-1}{r_0}.\label{Trefrac}
 \end{equation}
This expression agrees with the findings of \cite{evans2018effects}. It can also be recovered from the limit $\gamma=\infty$ of the exact solution derived in \ref{sec:exact}, which takes the form of a sum of exponentials $\exp(\lambda x)$, where the $\lambda$s are the roots of the polynomial (\ref{charpolySt}). For $x=0$,  $t^+_1=A^+_0+A^+_2$
where $A^+_0$ and $A^+_2$ are given by Eqs. (\ref{A2p})-(\ref{A0p}). In the limit $\gamma\rightarrow\infty$, the roots take the simple form $\lambda_1\approx\gamma$, $\lambda_2\approx-\sqrt{r_0}$ and $\lambda_3\approx\sqrt{r_0}$ [see also the line before Eq. (\ref{eq:denslargegamma2})]. Therefore, the exponential terms with $\lambda_1$ become dominant, and one finds $A^+_2\simeq(r_0+r_1)/[\gamma (r_0)^{3/2}]\to 0$ and
 \begin{equation}
     A^+_0\simeq\frac{(r_0+r_1) e^{\sqrt{r_0}}-r_1}{r_0 r_1}
 \end{equation}
which is Eq. (\ref{Trefrac}).


We now turn to the case of a strength $\gamma$ of order unity and $r_0\ll r_1\ll 1$.
In the expression for the roots given by Eq. (\ref{lambdas})-(\ref{theta}), we set $r_0=0$ and expand at first order in $r_1$, 
\begin{equation}
    \lambda_1\simeq\frac{r_1}{\gamma}+\gamma,\ \lambda_2\simeq-\frac{r_1}{\gamma},\  \lambda_3\simeq0.\label{eq:uneqs2}
\end{equation}
Replacing these values into Eqs. (\ref{A0p})-(\ref{A2p}) and keeping $r_0\neq0$ elsewhere, one obtains
\begin{equation}
  t_1\simeq \frac{r_0+r_1}{\gamma  r_0}+\frac{\gamma ^2\left(\gamma ^2+2 r_1\right)}{r_0 r_1 \left(\gamma ^2+r_1\right)}\left(\frac{r_0+r_1-\left(2 r_0+r_1\right) e^{\frac{r_1}{\gamma }}}{\gamma ^2+2 r_1
   e^{\gamma +\frac{2 r_1}{\gamma }}}+\frac{r_0}{\gamma ^2+2 r_1}\right)\label{t1approx1gam}.
   \end{equation}
Since the terms proportional to $1/r_0$ are dominant compared to those proportional to $1/r_1$, Eq. (\ref{t1approx1gam}) further simplifies to
   \begin{equation}
       t_1\simeq\frac{1}{r_0}\left[\frac{r_1}{\gamma }-\frac{\gamma ^2 \left(e^{\frac{r_1}{\gamma }}-1\right) \left(\gamma ^2+2 r_1\right)}{\left(\gamma ^2+r_1\right) \left(\gamma ^2+2 r_1 e^{\gamma
   +\frac{2 r_1}{\gamma }}\right)}\right],\label{secondApproxT1mu}
   \end{equation}
\begin{figure}[t]
    \centering
    \includegraphics[width=.6\textwidth]{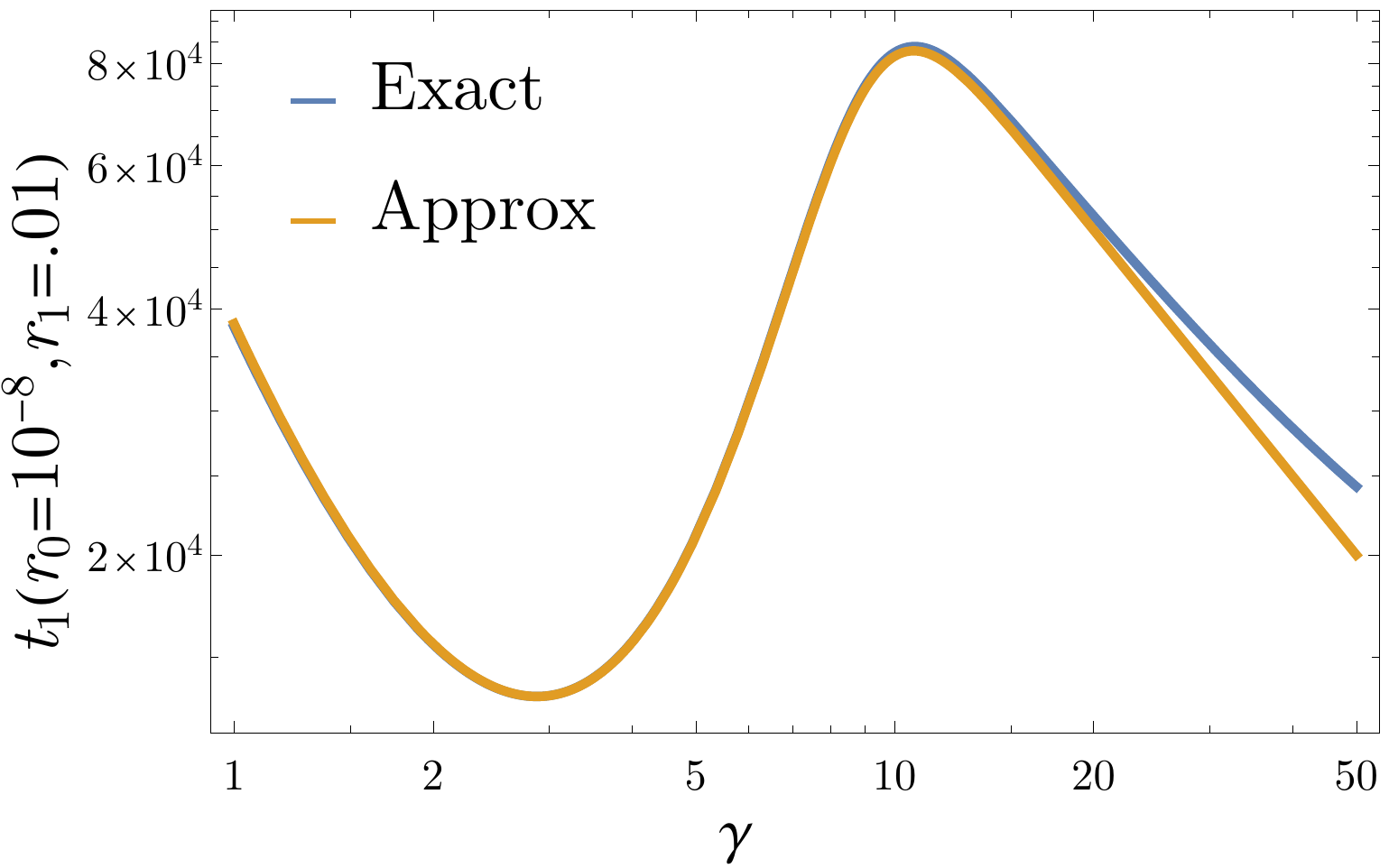}
    \caption{$t_1$ with $x=0$ as a function of $\gamma$ for $r_0=10^{-8}$ and $r_1=10^{-2}$. The exact solution (blue line) is compared with the approximation Eq. (\ref{secondApproxT1mu}) (orange line).}
    \label{fig:t1muapprox}
\end{figure}   
and one expects $t_1\gg1$. As illustrated in Figure \ref{fig:t1muapprox}, Eq. (\ref{secondApproxT1mu}) agrees very well with the exact solution at intermediate $\gamma$, where it actually presents a local minimum. We next expand the expression (\ref{secondApproxT1mu}) at small $r_1$, assuming $r_1\ll\gamma$, $r_1\ll\gamma^2$ and $r_1 e^{\gamma}/\gamma^2\ll 1$; these three inequalities are fulfilled if $\gamma$ is $O(1)$ and $r_1\ll 1$. The first non-zero term turns out to be of second order in $r_1$ and
  \begin{equation}
      t_1\simeq\frac{r_1^2}{r_0}\left(\frac{4 e^{\gamma }-\gamma-2}{2 \gamma ^3}\right),
\label{thirdApproxT1mu}
  \end{equation}
which is the result (\ref{t1smallr0r1}). At fixed rates, this expression is non-monotonic with $\gamma$ and reaches a minimum at a value independent of $(r_0,r_1)$, given by the root of $2 e^{\gamma } (\gamma -3)+\gamma +3=0$. A unique solution $\gamma_{min}=2.827641...$ is found, and the local minimum of the MFPT reads
 \begin{equation}
      t_1(\gamma_{min})\simeq 1.388733...\frac{r_1^2}{r_0}.
  \end{equation}
One can then compare this expression with the MFPT at $\gamma=\infty$, and deduce the transition line separating $\gamma^{\star}=\infty$ from $\gamma^{\star}<\infty$. 
Let us assume $t_1\gg1$ or $r_1\gg\sqrt{r_0}$ near the transition (an assumption to be verified {\it a posteriori}). Eq. (\ref{Trefrac}) reduces to $t_1(\gamma=\infty)\simeq 1/\sqrt{r_0}$ and
  \begin{equation}
      \frac{t_1(\gamma_{min})}{t_1(\gamma=\infty)}\simeq1.388733...\frac{r_1^2}{\sqrt{r_0}},\ {\rm for}\ \sqrt{r_0}\ll r_1\ll 1.
  \end{equation}
At the transition this ratio is exactly $1$, which gives the switch-off rate:
\begin{equation} r_1^c(r_0)=0.848575...r_0^{1/4}.\label{critline}
\end{equation}
One checks that $r_1^c(r_0)\gg\sqrt{r_0}$ in Eq. (\ref{critline}).
Summarizing, given $r_0\ll 1$, if one chooses a rate $r_1$ below $r_1^c(r_0)$, the optimal potential strength will be $\gamma^\star\simeq 2.827641$, while above this point the local minimum becomes metastable and  $\gamma^\star=\infty$.

\section{Conclusions}
We have studied the motion of a one-dimensional Brownian particle in an intermittent, symmetric confining potential. This problem extends well-known models of diffusion with stochastic resetting to the origin, which are recovered in some limits. Its implementation in experiments employing optical tweezers is possible, and would not require specific micro-manipulations such as bringing the particle back to the origin. We have shown that the presence of the potential, although intermittent in time, always leads to a non-equilibrium steady state and renders the mean first passage time to a target finite. The MFPT can be further minimized with respect to the re-scaled switch-on ($r_0$) and switch-off ($r_1$) rates.  Since the problem involves two rates instead of one as usual in resetting processes, we have found a rather rich phenomenology.

If the particle starts at the potential minimum, below a critical potential strength, the protocol that minimizes the MFPT to a target located at a certain distance consists in always keeping the potential on ($r_1=0$), whereas for stronger potentials the two optimal rates are non-zero. In the latter case, the dynamics are fully out-of-equilibrium and the trajectories a succession of free and biased diffusive phases. The transition in the optimal rates is continuous or discontinuous at the critical potential strength, depending on whether the potential is initially on or off. 
Above threshold, the optimal MFPT is nearly always lower than the lowest Kramers' time, a result that generalizes previous comparisons between the efficiency of equilibrium and non-equilibrium searches \cite{evans2013optimal,giuggioli2019comparison}. 
Likewise, when one seeks to minimize the MFPT with respect to the potential strength at fixed rates, another discontinuous transition occurs and a phase diagram can be drawn in the $(r_0,r_1)$-plane.

Near the continuous transition mentioned above, a perturbative theory allows to decompose the MFPT into an equilibrium part (which depends only on the potential strength) and a rate dependent part. The latter is non-monotonous with respect to $r_0$ and reaches a  minimum at a value $\gg 1$ which contrasts with the optimal re-scaled rates of order unity found in usual resetting problems. Importantly, this contribution to the MFPT can take negative values above the critical strength, a regime where minimization leads to non-trivial rates in a way analogous to a second order Ginzburg-Landau transition.

In the vicinity of the discontinuous transitions, we have unveiled metastable behaviors as the rates or the strength are varied. Metastability in a resetting process was recently reported and confirmed experimentally with optical traps in \cite{besga2020optimal} using a quite different setup and protocol. Hence, this new property is likely to be generic in resetting processes. 

Our study shows that incorporating physical constraints into idealized resetting models (which often assume instantaneous relocations, for instance) can lead to new phenomena. Our calculations were performed with a linear potential, but we expect similar conclusions with harmonic or other confining potentials, at least qualitatively. The distribution of the first passage times remains unknown and could be investigated through the calculation of higher moments.
A first-passage distribution with spectacular spikes was found and explained theoretically in optical trap experiments with periodic resetting \cite{besga2020optimal}. Resetting protocols at periodic times \cite{pal2016diffusion,chechkin2018random,nagar2016diffusion} that act on a potential deserve further study.

\ack 
This work was supported by a grant Investissements d'Avenir from LabEx PALM (ANR-10-LABX-0039-PALM). DB thanks CNRS for support and the Laboratoire de Physique Th\'eorique et Mod\`eles Statistiques for hospitality. GMV thanks CONACYT (Mexico) for a scholarship support.

\appendix
 
 \section{Exact expressions for the MFPTs}\label{sec:exact}
We derive the complete solution which is used throughout this study.
Let us start from Eq. (\ref{difT12}) for $t_1^{+}(x)$. It admits the inhomogeneous solution $x(r_1+r_0)/(r_0\gamma)$, whereas the homogeneous anzats $e^{\lambda x}$ leads to an equation for $\lambda$ which is given by Eq. (\ref{charpolySt}). The admissible roots are $\lambda_0=0$ and $\lambda_2$ [see Eq. (\ref{eq:uneqs})], as they avoid exponential divergence at $x\rightarrow+\infty$:
\begin{equation}
    t^+_1(x)=A^+_0+A^+_2 e^{\lambda_2 x}+\frac{r_1+r_0}{r_0\gamma}x,\label{completet1p}
\end{equation}
with $A_{0,2}^+$ two constants. Similarly, in the interval $-1\le x<0$, the inhomogeneous solution of Eq. (\ref{difT1m2}) for $t_1^{-}(x)$ is $-x(r_1+r_0)/(r_0\gamma)$, and the homogeneous solutions take the form $e^{-\lambda x}$, with $\lambda$ a root of the same polynomial (\ref{charpolySt}). Therefore
\begin{equation}
    t^-_1(x)=A_0^{-}+\sum_{k=1}^{3}A^{-}_k e^{-\lambda_k x}-\frac{r_1+r_0}{r_0\gamma}x.\label{completet1m}
\end{equation}
Thus we have 6 constants to determine, from the 6 boundary conditions (\ref{BC1})-(\ref{BC3}).
Using Eqs. (\ref{eq:relt0t1})-(\ref{eq:relt0t1m}), $t_0$ is deduced from $t_1$ on each side as:
\begin{equation}
    t_0^+(x)=A_0^+ +\frac{1}{r_0}+A_2^+\left[1+\frac{\lambda_2(\lambda_1+\lambda_3)}{r_1} \right]e^{\lambda_2x}+\frac{r_1+r_0}{r_0\gamma}x
\end{equation}
and
\begin{equation}
    t_0^-(x)=A_0^-+\frac{1}{r_0}+\sum_{k=1}^3A_k^-\left[1+\frac{\lambda_k(\lambda_i+\lambda_j)}{r_1}\right]e^{-\lambda_k x}-\frac{r_1+r_0}{r_0\gamma}x,
\end{equation}
where we have used the identity  $\lambda_1+\lambda_2+\lambda_3=\gamma$ and  where $i,j$ represent the two indices different from $k$.  The boundary conditions (\ref{BC1})-(\ref{BC3}) lead to the system:
\begin{equation}
 \fl   \left(\begin{array}{cccccc}
    1 & -1 & -1 & -1 & 1 & -1 \\
    \lambda_2(\lambda_1+\lambda_3) & -\lambda_3(\lambda_1+\lambda_2) & -\lambda_2(\lambda_1+\lambda_3) & -\lambda_1(\lambda_2+\lambda_3) & 0 & 0 \\
    \lambda_2 & \lambda_3 & \lambda_2 & \lambda_1 & 0 & 0 \\
    \lambda^2_2(\lambda_1+\lambda_3) & \lambda^2_3(\lambda_1+\lambda_2) & \lambda^2_2(\lambda_1+\lambda_3) & \lambda^2_1(\lambda_2+\lambda_3) & 0 & 0 \\
    0 & e^{\lambda_3} & e^{\lambda_2} & e^{\lambda_1} & 0 & 1 \\
    0 & \lambda_3(\lambda_1+\lambda_2)e^{\lambda_3} & \lambda_2(\lambda_1+\lambda_3)e^{\lambda_2} & \lambda_1(\lambda_2+\lambda_3)e^{\lambda_1} & 0 & 0 
    \end{array}\right)\left(\begin{array}{c}
    A_2^+\\
    A_3^-\\
    A_2^-\\
    A_1^-\\
    A_0^+\\
    A_0^-
    \end{array}\right)=\left(\begin{array}{c}
    0\\
    0\\
    -2\frac{r_1+r_0}{r_0\gamma}\\
    0\\
    -\frac{r_1+r_0}{r_0\gamma}\\
    -\frac{r_1}{r_0}
    \end{array}\right).\label{solutionmatrix}
\end{equation}
The MFPTs we are interested in are:
\begin{eqnarray}
     t^+_1(x=0)=A^+_0+A^+_2,\label{solt1}\\
     t_0^+(x=0)=A_0^+ +\frac{1}{r_0}+A_2^+\left[1+\frac{\lambda_2(\lambda_1+\lambda_3)}{r_1}\right].\label{solt0}
\end{eqnarray}
For convenience we use the notation
\begin{eqnarray}
     \Lambda^-_1&=\lambda_1-\lambda_2,\qquad\qquad\Lambda^+_1= \lambda_1+\lambda_2,\\
    \Lambda^-_2&=\lambda_2-\lambda_3,\qquad\qquad\Lambda^+_2=\lambda_2+\lambda_3,\\
     \Lambda^-_3&=\lambda_3-\lambda_1,\qquad\qquad\Lambda^+_3=\lambda_3+\lambda_1.
 \end{eqnarray}
Solving the system (\ref{solutionmatrix}) yields
\begin{equation}
A_2^+=
    -\frac{\Lambda^-_1e^{\lambda _3}+\Lambda^-_3e^{\lambda _2} +\Lambda^-_2e^{\lambda _1}+\frac{r_1\gamma}{2(r_0+r_1)}\frac{  \Lambda^-_1\Lambda^-_2\Lambda^-_3}{ \Lambda^+_1 \Lambda^+_2\Lambda^+_3}}{  \frac{r_0\gamma \lambda _2}{r_0+r_1}\left(\frac{\lambda _3\Lambda^-_1}{\Lambda^+_2}e^{\lambda _3}+\frac{1}{2}\Lambda^-_3e^{\lambda _2}+\frac{\lambda _1 \Lambda^-_2}{\Lambda^+_1}e^{\lambda _1}\right)},\label{A2p}
\end{equation}
and
\begin{equation}
 \fl \eqalign{
 A_0^+=&-\left(\frac{r_0+r_1}{\gamma  r_0}\right)\Bigg[\frac{2 \lambda _2^2 e^{\Lambda^+ _3} \Lambda^+ _3 \Lambda^-_3}{\lambda _1 \lambda _3 \Lambda^+ _1 \Lambda^+ _2}+\frac{\lambda _3
   e^{\Lambda^+_1} \Lambda^-_1}{\lambda _1 \lambda _2}+\frac{e^{\lambda _3} \Lambda^-_1 \left(\left(\frac{2}{\lambda _1}+1\right) \lambda _3-\frac{\gamma  r_1}{\Lambda^+ _1
   \left(r_0+r_1\right)}\right)}{\Lambda^+ _2}\\
   &+\frac{e^{\lambda _1} \Lambda^-_2 \left(\lambda _1 \left(\frac{2}{\lambda
   _3}+1\right)-\frac{\gamma  r_1}{\Lambda^+ _2 \left(r_0+r_1\right)}\right)}{\Lambda^+ _1}+\frac{e^{\lambda _2} \Lambda^-_3
   \left(-\frac{2 \left((\gamma +1) r_0+r_1\right)}{\lambda _1 \lambda _3}-\frac{\gamma  r_1}{\Lambda^+_3 \left(r_0+r_1\right)}-4\right)}{2
   \lambda _2}\\
   &+\frac{\lambda _1 e^{\Lambda^+ _2} \Lambda^-_2}{\lambda _3 \lambda
   _2}-\frac{\gamma  \Lambda^-_1 \Lambda^-_2 \Lambda^-_3 r_1}{\lambda _2 \Lambda^+ _1 \Lambda^+ _2 \Lambda^+ _3
   \left(r_0+r_1\right)}\Bigg]\left(\frac{e^{\lambda _3} \lambda _3 \Lambda^-_1}{\Lambda^+ _2}+\frac{ e^{\lambda _1} \lambda _1 \Lambda^-_2}{\Lambda ^+_1}+\frac{1}{2}e^{\lambda _2} \Lambda^-_3\right)^{-1}.\label{A0p}
   }
\end{equation}

\section{Expressions for $t_1^{(1)}$ and $t_1^{(2)}$}\label{appendixt2}
The solution $t_1^{(1)}(x)$ for $x\in[0,\infty)$ is given by
    \begin{equation}
    \fl \eqalign{
    t_1^{(1)}(x,+)=&\left(\frac{\sqrt{r_0}e^{-\sqrt{r_0}} \left(1-2 e^{\gamma }-\frac{\gamma ^2}{r_0}\right)+\gamma
   +\sqrt{r_0}}{\left(\sqrt{r_0}-\gamma\right)\left(\sqrt{r_0}+\gamma
   \right)^2 }+\frac{e^{-2 \sqrt{r_0}}}{\left(\sqrt{r_0}+\gamma
   \right)^2}\right)\left(e^{-\sqrt{r_0}x}-1\right)\\
   &-\gamma\frac{\partial t_1^{+(0)}(x)}{\partial \gamma}+\frac{r_0\gamma^2\left[  \frac{4 e^{\gamma }}{\sqrt{r_0}\gamma}- \left(\frac{2 e^{\gamma }-1}{\gamma}+\frac{\gamma}{r_0}\right) \left(\frac{2 e^{\gamma }-1}{\gamma}+\frac{1}{\sqrt{r_0}}\right)e^{-\sqrt{r_0}} \right]}{\left(r_0-\gamma ^2\right){}^2}\\
   &+\frac{\gamma \sqrt{r_0}\left(\frac{2 e^{\gamma }-1}{\gamma }-\frac{2 e^{\sqrt{r_0}}-1}{\sqrt{r_0}}\right) e^{-2 \sqrt{r_0}}}{\left(\gamma +\sqrt{r_0}\right) \left(r_0-\gamma ^2\right)}-\frac{2 e^{\gamma } (\gamma -1)+\frac{\gamma }{\sqrt{r_0}}+2}{r_0-\gamma ^2},}
    \end{equation}
and the solution in the interval $x\in[-1,0]$ is:
\begin{equation}
\fl \eqalign{
    t_1^{(1)}(x,-)=&\left(\frac{4 r_0 e^{\gamma -\sqrt{r_0}}}{\left(r_0-\gamma ^2\right){}^2}-\frac{2 \left(e^{-\sqrt{r_0}}+1\right)\left(\gamma 
   +\sqrt{r_0}e^{-\sqrt{r_0}}\right)}{\left(\sqrt{r_0}+\gamma\right)\left(r_0-\gamma ^2\right)}\right)\left(e^{-\gamma x}-1\right)-\frac{e^{\sqrt{r_0} x}-1}{\left(\gamma +\sqrt{r_0}\right){}^2}\\
    &-\frac{e^{-2 \sqrt{r_0}} \sqrt{r_0}\left(\frac{\gamma ^2 }{r_0}e^{\sqrt{r_0}}+\frac{\gamma}{\sqrt{r_0}} +1-\left(1-2 e^{\gamma }\right) e^{\sqrt{r_0}}\right)\left(e^{-\sqrt{r_0}x }-1\right)}{\left(\sqrt{r_0}-\gamma \right){}^2 \left(\gamma +\sqrt{r_0}\right)}-\frac{2 \gamma  x
   e^{-x\gamma }}{r_0-\gamma ^2}\\
    &-\gamma\frac{\partial t_1^{-(0)}(x)}{\partial \gamma}+\frac{r_0\gamma^2\left[  \frac{4 e^{\gamma }}{\sqrt{r_0}\gamma}- \left(\frac{2 e^{\gamma }-1}{\gamma}+\frac{\gamma}{r_0}\right) \left(\frac{2 e^{\gamma }-1}{\gamma}+\frac{1}{\sqrt{r_0}}\right)e^{-\sqrt{r_0}} \right]}{\left(r_0-\gamma ^2\right){}^2}\\
   &+\frac{\gamma \sqrt{r_0}\left(\frac{2 e^{\gamma }-1}{\gamma }-\frac{2 e^{\sqrt{r_0}}-1}{\sqrt{r_0}}\right) e^{-2 \sqrt{r_0}}}{\left(\gamma +\sqrt{r_0}\right) \left(r_0-\gamma ^2\right)}-\frac{2 e^{\gamma } (\gamma -1)+\frac{\gamma }{\sqrt{r_0}}+2}{r_0-\gamma ^2}.
   }
\end{equation}

For the coefficient of the term of order $\epsilon^2$, at $x=0$, we obtain:
\begin{equation}
\fl \eqalign{
    t^{(2)}_1(\gamma,r_0)=&\frac{\gamma^5  r_0^2e^{-3 \sqrt{r_0}} \left(\frac{1}{r_0^2}+2 \frac{e^{\gamma }-2}{r_0^{3/2}\gamma}+4 \frac{e^{\gamma }+1}{ r_0\gamma ^2}+2\frac{ e^{\gamma } \left(6 e^{\gamma }-5\right)}{\sqrt{r_0}\gamma^3} -\frac{\left(1-2 e^{\gamma }\right)^2}{\gamma^4}\right)}{2 \left(\gamma
   -\sqrt{r_0}\right){}^3 \left(\gamma +\sqrt{r_0}\right){}^4}+\frac{e^{-4 \sqrt{r_0}}r_0 \gamma ^2\left(\frac{1 -2 e^{\gamma }}{\gamma}-\frac{1}{\sqrt{r_0}}\right)}{2 \left(\sqrt{r_0}-\gamma \right)
   \left(\gamma +\sqrt{r_0}\right){}^4}\\
   &+\frac{r_0^2\gamma^2 e^{-2 \sqrt{r_0}}}{\left(r_0-\gamma ^2\right){}^2} \Bigg(\frac{4 e^{2\gamma} \left(\frac{\gamma ^2}{r_0}+\frac{\gamma }{\sqrt{r_0}}-2\right)+8 e^{3 \gamma }}{ \left(r_0-\gamma ^2\right){}^2}-2 e^{\gamma }\left(\frac{3}{r_0}+\frac{2}{\gamma}+\frac{\gamma -1}{\gamma\sqrt{r_0}}+\frac{\sqrt{r_0}+1}{\gamma ^2}\right)\\
   &- \frac{2\gamma^2}{r_0^2}-\frac{\gamma}{2r_0^{3/2}} +\frac{3}{r_0}-\frac{2 \gamma +7} {2\gamma\sqrt{r_0}}+\frac{1+\sqrt{r_0}}{\gamma^2}\Bigg)+\frac{\gamma^6 r_0^3 e^{-\sqrt{r_0}}}{ \left(r_0-\gamma ^2\right){}^4}\Bigg(\frac{\left(1-2 e^{\gamma }\right)^2 \left(\sqrt{r_0}+2\right)}{2\gamma ^6}\\
   &+\frac{e^{\gamma } \left(9 \sinh
   (\gamma )+11 \cosh (\gamma )-5+\frac{7-4 \sinh (\gamma )}{\sqrt{r_0} }\right)}{\gamma ^5 }-\frac{\sqrt{r_0}-4}{2r_0^3}+\frac{3 e^{\gamma
   }+1}{\gamma  r_0^2}-\frac{5 e^{\gamma }+2}{\gamma  r_0^{5/2}}\\
   &-\frac{2 \left(e^{\gamma } \left(5 e^{\gamma }
   \left(\sqrt{r_0}+1\right)-4 \sqrt{r_0}-5\right)-6\right)}{\gamma ^3 r_0^{3/2}}-\frac{2 e^{\gamma }\left(\frac{2 e^{\gamma }-1}{\sqrt{r_0}}+2\right)-\frac{7}{2} \left(1-\frac{2}{\sqrt{r_0}}\right)}{\gamma ^2 \sqrt{r_0}}\\
   &-\frac{2 e^{\gamma } \left(e^{\gamma }
   \left(\sqrt{r_0}-10\right)-3 \sqrt{r_0}+5\right)+\frac{7}{2} \sqrt{r_0}+8}{\gamma ^4 r_0}\Bigg)+\frac{\gamma^6 r_0^4}{\left(r_0-\gamma ^2\right){}^4}\Bigg(\frac{8 e^{\gamma }+1}{2\gamma  r_0^{7/2}}\\
   &-\frac{2 \left(e^{\gamma } ((\gamma -5) \gamma +5)-5\right)}{\gamma ^6
   r_0}-\frac{4 e^{\gamma } (2 \gamma +5)-5}{2\gamma ^5 r_0^{3/2}}+\frac{2 e^{\gamma } ((\gamma -10) \gamma +10)-21}{2\gamma ^4 r_0^2}\\
   &+\frac{
   2 e^{\gamma } (2 \gamma -7)-3}{\gamma ^3 r_0^{5/2}}+\frac{2 e^{\gamma } (\gamma +1)+3}{\gamma ^2
   r_0^3}+\frac{e^{\gamma } ((\gamma -2) \gamma +2)-2}{\gamma ^8}-\frac{1}{2r_0^4}\Bigg).}\label{t12}
\end{equation}



\section*{References}
\bibliographystyle{iopart-num}
\bibliography{Biblio}
\end{document}